%% file: Thesis.tex
\documentclass[twoside,12pt]{report}
\usepackage{graphics}
\usepackage{epsfig,subfigure}
\usepackage[centertags]{amsmath}
\usepackage{amsfonts}
\usepackage{amssymb}
\usepackage{amsthm}
\usepackage{newlfont}
\usepackage{XThesis} 
\usepackage{XTocinc} 
\newlength{\defbaselineskip}
\newcommand{\setlinespacing}[1]%
           {\setlength{\baselineskip}{#1 \defbaselineskip}}

\newcommand{\A}{{\cal A}}

\newcommand{\dC}{\mathbb C}

\newcommand{\dR}{\mathbb R}
\newcommand{\dS}{\mathbb S}

\newcommand{\Milne}{\mathcal{M}}
\newcommand{\sing}{\mathcal{S}}
\newcommand{\dZ}{\mathbb Z}
\newcommand{\up}{\uparrow}
\newcommand{\down}{\downarrow}
\newcommand{\var}{\varepsilon}

\newcommand{\Ola}{\Omega_{\check{\lambda}}}
\numberwithin{equation}{section}

\begin{document}



\dedicate{To my parents}

\nolistoftables \nolistoffigures 
 \phd



\title{Modelling cosmological singularity \\
with compactified Milne space}

\author{Przemys\l aw Ma\l kiewicz}




{
\typeout{:?0000} 
\beforepreface
\typeout{:?1111} 
}
{
\typeout{Acknowledgements}
\include{ACK}
}
\afterpreface
\def\baselinestretch{1}
\setlinespacing{1.66}
{ \typeout{Introduction}
\include{T0}
}
\setlinespacing{1.66}

\include{T1}
\include{T2}
\include{T3}

\appendix
\include{T4}

\setlinespacing{1.44}
\bibliographystyle{plain}
\bibliography{XBib}
\end{document}

%% file: ACK.tex

\prefacesection{Acknowledgements}
\def\baselinestretch{1.0}
\setlinespacing{1.15}

I would like to thank doc. dr hab. W\l odzimierz Piechocki, my
supervisor, who introduced me to the problems and methods of
Quantum Cosmology and with whom I shared a pleasure of joint
scientific investigations.

\noindent I am grateful to prof. A. Sym and dr M. Nieszporski,
whose kind support helped me to take up physics more seriously.

\noindent This work has been supported by the Polish Ministry of
Science and Higher Education Grant  NN 202 0542 33.
\\

\noindent
Warsaw \hfill Przemys\l aw Ma\l kiewicz\\
May 30, 2009


%% file: T0.tex

\nonumchapter{Introduction}

Presently available cosmological data  suggest that the Universe
emerged from a state with extremely high density of physical
fields. It is called the cosmological singularity. The data also
indicate that known forms of energy and matter comprise only $4\%$
of the makeup of the Universe. The remaining $96\%$ is unknown,
called `dark', but its existence is needed to explain the
evolution of the Universe \cite{Spergel:2003cb,Bahcall:1999xn}.
The dark matter, DM, contributes $22\%$ of the mean density. It is
introduced to explain the observed dynamics of galaxies and
clusters of galaxies. The dark energy, DE, comprises $74\%$ of the
density and is responsible for the observed accelerating
expansion. These data mean that we know almost nothing about the
dominant components of the Universe!

Understanding the nature and the abundance of the DE and DM within
the standard  model of cosmology, SMC, has  difficulties
\cite{GDS,NGT}. These difficulties have led many physicists to
seek anthropic explanations which, unfortunately, have little
predictive power. However, there exist promising models based on
the idea of a cyclic evolution of the Universe. There are two main
developments based on such an idea: (i) resulting from application
of loop quantum gravity \cite{AAJL,Rov,TT} to quantization of FRW
type Universes, and (ii) inspired by string/M theory
\cite{Khoury:2001bz}, the so called cyclic model of the Universe,
CMU \cite{Steinhardt:2001st,Steinhardt:2001vw}.

The loop quantum cosmology, LQC, shows that the classical
cosmological singularity does not occur due to the loop geometry.
The Big-Bang of the SMC model is replaced by the Big-Bounce
\cite{12,11,14,13}. However, at the present state of development,
the LQC is unable to explain the origin of DE and DM.

An alternative model has been proposed by Steinhardt and Turok
(ST) \cite{Steinhardt:2001st,Steinhardt:2001vw,Steinhardt:2004gk}.
The ST model has been inspired by string/M  theories
\cite{Khoury:2001bz}. In its simplest version it assumes that the
spacetime can be modelled by the higher dimensional compactified
Milne space, $\Milne_C$. The most developed model
\cite{Steinhardt:2001vw,Steinhardt:2001st} is one in which
spacetime is  assumed to be the five dimensional compactified
Milne space. In this model the Universe has a form of two
4-dimensional branes separated by a distance which changes
periodically its length from zero to some finite value. The
Universe changes periodically its dimensionality from five to
four, which leads to the evolution of the Universe of the
Big-Crunch / Big-Bang type. This model tries to explain the
observed properties of the Universe as the result of interaction
of `our' brane with the other one. The attractiveness of the ST
model is that it potentially provides a complete scenario of the
evolution of the universe, one in which the DE and DM play a key
role in both the past and the future. The ST model
\textit{requires} DE for its consistency, whereas in the standard
model, DE is introduced in a totally \textit{ad hoc} manner.
Demerits of the ST model  are extensively discussed in
\cite{Linde:2002ws}. Response to the criticisms of
\cite{Linde:2002ws} can be found in \cite{NGT}.

The mathematical structure and self-consistency of the ST model
has yet not been fully tested and understood. Such task presents a
serious mathematical challenge. It is the subject of the Thesis.

The CMU model has in each of its cycles a quantum phase including
the cosmological singularity, CS. The CS plays key role because it
joins each two consecutive classical phases.  Understanding the
nature of the CS has  primary importance for the CMU model.  Each
CS consists of contraction and expansion phases. \emph{A
physically correct model of the CS, within the framework of
string/M theory, should be able to describe  propagation of a
p-brane, i.e. an elementary object like a particle, string and
membrane, from the pre-singularity to post-singularity epoch}.
This is the most elementary, and fundamental, criterion that
should be satisfied. It presents a new criterion for testing the
CMU model. Hitherto, most research has focussed on the evolution
of scalar perturbations through the CS.

Successful quantization of the dynamics of p-brane will mean that
the $\Milne_C$ space is a promising candidate to model the
evolution of the Universe at the cosmological singularity. Thus,
it could be further used in advanced numerical calculations to
explain  the data of observational cosmology. Failure in
quantization may mean that the CS should be modelled by a
spacetime more sophisticated than the $\Milne_C$ space.


\begin{figure}[h]
\center
\includegraphics[width=0.3\textwidth]{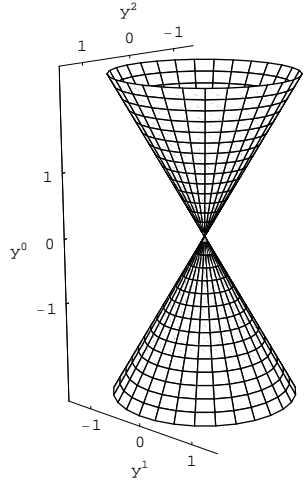}
\caption{Compactified 2d Milne space embedded in 3d Minkowski
space.}
\end{figure}

The figure $1$ shows the two dimensional $\Milne_C$ space embedded in the three dimensional Minkowski space. It can be specified by the
following isometric embedding
\begin{equation}\label{emb}
    y^0(t,\theta) = t\sqrt{1+r^2},~~~~y^1(t,\theta) =
    rt\sin(\theta/r),~~~~y^2(t,\theta) = rt\cos(\theta/r),
\end{equation}
where $ (t,\theta)\in \dR^1 \times \dS^1 $ and $ 0<r \in\dR^1 $ is
a constant labelling compactifications . One has
\begin{equation}\label{stoz}
    \frac{r^2}{1+r^2}(y^0)^2 - (y^1)^2- (y^2)^2 =0.
\end{equation}
Eq. (\ref{stoz}) presents two cones with a common vertex at
$\:(y^0,y^1,y^2)= (0,0,0)$. The induced metric on (\ref{stoz})
reads
\begin{equation}\label{line1}
    ds^2 =  - dt^2 +t^2 d\theta^2 .
\end{equation}
Generalization of the 2-dimensional CM space to the $d+1$
dimensional spacetime has the form
\begin{equation}\label{line2}
ds^2 = -dt^2  +  t^2 d\theta^2 + \delta_{kl}~dx^k dx^l ,
\end{equation}
where $t,x^k \in \mathbb{R}^1,~\theta\in \mathbb{S}^1~(k=
2,\ldots, d)$.

One term in the metric (\ref{line2}) disappears/appears at $t=0$,
thus the $\Milne_C$ space may be used to model the
big-crunch/big-bang type singularity. Orbifolding  $\dS^1$ to the
segment gives a model of spacetime in the form of two orbifold
planes which collide and re-emerge at $t=0$. Such a model of
spacetime was used in
\cite{Khoury:2001bz,Steinhardt:2001st,Steinhardt:2001vw}. Our
results apply to both choices  of topology of the compact
dimension.

The $\Milne_C$ space is an orbifold due to the vertex at $t=0$.
The Riemann tensor components equal $0$ for $t\neq 0$. The
singularity at $t=0$ is of removable type: any time-like geodesic
with $t<0$ can be extended to some time-like geodesic with $t>0$.
However, the extension cannot be unique due to the Cauchy problem
at $t=0$ for the geodesic equation  (the compact dimension shrinks
away and reappears at $t=0$).


%% file: T1.tex
\def\baselinestretch{1}

\chapter{Classical dynamics of extended objects}

\def\baselinestretch{1.66}


In this chapter we consider classical dynamics of $p$-brane propagating in background spacetime. We formulate it in terms of both Lagrangian and
Hamiltonian. The formulations admit gauge symmetry: the action is invariant with respect to diffeomorphisms of $p$-brane's world-sheet and the
Hamiltonian is a sum of first-class constraints. Next we specialize the formalism to the case the embedding spacetime is the compactified Milne
space, $\Milne_C$, and analyze classical propagation of extended objects as well as prepare formalism for canonical quantization.

\smallskip

\goodbreak
\section{Lagrangian formalism}

A $p$-brane is a $p$-dimensional object, which traces out a
$p+1$-dimensional surface, called a $p$-brane's world-sheet, in
the embedding spacetime as it propagates. Both the embedding
spacetime and the world-sheet are assumed to be locally
Lorentzian.

\bigskip
\goodbreak

The Nambu-Goto action is a $p+1$-volume of the p-brane world-sheet
and reads:
\begin{equation}\label{e:nambu}
S_{N-G}=-\mu_p\int
\sqrt{|det(g^{IND}_{ab})|}~d^{p+1}\sigma=-\mu_p\int
\sqrt{-det(\partial_aX^{\alpha}\partial_bX^{\beta}{g}_{\alpha\beta})}~d^{p+1}\sigma,
\end{equation}
where $\mu_p$ is a mass per unit $p+1$-volume,
$(\sigma^a)\equiv(\sigma^0,\sigma^1,\ldots,\sigma^p)$ are
$p$-brane world-sheet coordinates, $g^{IND}_{ab}$ is an induced
metric on the world-sheet, $~(X^{\alpha})\equiv (X^\mu,
\Theta)\equiv (T,X^k,\Theta)\equiv (T,X^1,\ldots,X^{d-1},\Theta)$
are the embedding functions of a $p$-brane, i.e. $X^{\alpha} =
X^{\alpha}(\sigma^0,\ldots,\sigma^p$), in $d+1$ dimensional
background spacetime with metric ${g}_{\alpha\beta}$. As a subcase
for $p=0$ the formula (\ref{e:nambu}) includes the action of a
particle moving in a background spacetime. The least action
principle, i.e. $\delta S_{N-G}=0$, applied to (\ref{e:nambu})
leads to the following equations of motion:
\begin{eqnarray}\nonumber
\partial_a(\frac{\partial_bX^{\alpha}\partial_bX^{\beta}g_{\alpha\beta}}
{\sqrt{-det(\partial_aX^{\alpha}\partial_bX^{\beta}{g}_{\alpha\beta})}}
\partial_aX_{\mu}-
\frac{\partial_aX^{\alpha}\partial_bX^{\beta}g_{\alpha\beta}}
{\sqrt{-det(\partial_aX^{\alpha}\partial_bX^{\beta}{g}_{\alpha\beta})}}
\partial_bX_{\mu})\\ \label{e:nambueom}
-\frac{(\partial_aX^{\alpha}\partial_aX^{\beta}g_{\alpha\beta})\partial_bX^{\alpha}
\partial_bX^{\beta}-(\partial_aX^{\alpha}\partial_bX^{\beta}g_{\alpha\beta})
\partial_aX^{\alpha}\partial_bX^{\beta}
}{2\sqrt{-det(\partial_aX^{\alpha}\partial_bX^{\beta}{g}_{\alpha\beta})}}g_{\alpha\beta
,\mu}=0 .
\end{eqnarray}
The above equations (\ref{e:nambueom}) are undetermined (not only
because of unspecified initial/boundary conditions but) due to
freedom in the choice of parameters $(\sigma^a)$ (for
$a=0,\dots,p$) as consequence of re-parametrization invariance of
the action (\ref{e:nambu}). A convenient setting for gauge fixing
is the Polyakov action.

\medskip

The Polyakov action for a test $p$-brane   embedded in a
background spacetime with  metric $g_{\alpha\beta}$ has the form
\begin{equation}\label{e:polyakov}
    S_P= -\frac{1}{2}\mu_p \int d^{p+1}\sigma
    \sqrt{-\gamma}\;\big(\gamma^{ab}\partial_a X^{\alpha} \partial_b
    X^{\beta}
    g_{\alpha\beta}-(p-1)\big),
\end{equation}
where $\gamma_{ab}$ is the $p$-brane world-sheet metric, $\gamma
:= det[\gamma_{ab}]$. The least action principle applied to
(\ref{e:polyakov}) produces the following equations of motion:
\begin{equation}\label{e:polyakoveom1}
    \partial_a(\sqrt{-\gamma}\gamma^{ab}\partial_bX_{\mu})=\frac{1}{2}\sqrt{-\gamma}
    \gamma^{ab}\partial_aX^{\alpha}\partial_bX^{\beta}g_{\alpha\beta,\mu},
\end{equation}
\begin{equation}\label{e:polyakoveom2}
 \partial_aX^{\alpha}\partial_bX^{\beta}~g_{\alpha\beta}-\frac{1}{2}\gamma_{ab}
 \gamma^{cd}\partial_cX^{\alpha}\partial_dX^{\beta}~g_{\alpha\beta}=0.
\end{equation}
The above equations are in full equivalence with the equations
(\ref{e:nambueom}). But in this case it is convenient to fix a
gauge by specifying the fields $\gamma_{ab}$ to some extent. For
example, in case of a string there are two ways of doing it:
\begin{enumerate}
    \item Partially fixed gauge: one sets the matrix
    $\sqrt{\gamma}\gamma^{ab}$ as functions of $(\sigma^a)$; afterwards there are still conformal
    isometries of the world-sheet allowed in this setting and the least action principle wrt fields
    $X^{\alpha}$ is still applicable.
    \item Fully fixed gauge: one sets lapse and shift function like in General Relativity; one fixes this gauge at the
    level of equations of motion.
\end{enumerate}

\smallskip

In the next section we will move to the Hamiltonian formalism,
which comes from applying a Legandre transormation to the
Nambu-Goto or Polyakov action.

\bigskip

\goodbreak

\def\baselinestretch{1.1}

\section{Hamiltonian formalism}\label{s:ham}

This section introduces Hamiltonian formalism with a brief review of Dirac's procedure for constrained systems. The constraints are phase space
functions that are gauge generators, i.e. they are manifestation of re-parametrization invariance of the corresponding action.

\def\baselinestretch{1.66}

\bigskip
\goodbreak

Let us denote a position-velocity space of a system by
$(q,\dot{q})$. Let us also assume that the Legendre transformation
$(q,\dot{q})\mapsto (q,p=\frac{\partial L}{\partial\dot{q}})$ is
singular, i.e. there exist relations of the form $\Phi_a(q,p)=0$.
The consistency condition requires:
\[ \{\Phi_a, H\} \approx 0,~~\{\Phi_a, \Phi_b\} \approx 0,\] where $H=p\dot{q}-L$, '$\approx$'
denotes equality holding on the surface $\Phi(q,p)=0$ and
$a,b=1,2,\dots$ The satisfaction of the above equation may require
introduction of new relations $\Upsilon_a(q,p)=0$, called
secondary constraints. One applies the consistency condition until
it produces no more new constraints. Now the constraints are
first-class, which means they close to a Poisson algebra (for more
details see \cite{PAM,MHT}).

Sometimes it is possible to reduce the number of conjugate pairs
by solving some of the constraints. This is called reduced phase
space formalism and it is used here.

It has been found  \cite{Turok:2004gb} that the total Hamiltonian,
$H_T$, corresponding to the action (\ref{e:nambu}) is the
following
\begin{equation}\label{ham}
H_T = \int d^p\sigma \mathcal{H}_T,~~~~\mathcal{H}_T := A C + A^i
C_i,~~~~~i=1,\ldots,p
\end{equation}
where $A=A(\sigma^a)$ and $A^i = A^i(\sigma^a)$ are any functions
of $p$-volume coordinates,
\begin{equation}\label{conC}
    C:=\Pi_{\alpha} \Pi_{\beta} g^{\alpha\beta} + \mu_p^2 \;det[\partial_a X^{\alpha} \partial_b
    X^{\beta} g_{\alpha\beta}]\approx 0,
\end{equation}
\begin{equation}\label{conCi}
    C_i := \partial_i X^{\alpha} \Pi_{\alpha} \approx 0,
\end{equation}
and where $\Pi_{\alpha}$ are the canonical momenta corresponding
to $X^{\alpha}$. Equations (\ref{conC}) and (\ref{conCi}) define
the first-class constraints of the system.

The Hamilton equations are
\begin{equation}\label{hameq}
    \dot{X}^{\alpha}\equiv\frac{\partial{X}^{\alpha}}{\partial\tau}=
    \{X^{\alpha},H_T \},~~~~~~\dot{\Pi}_{\alpha}\equiv\frac{\partial{\Pi}_{\alpha}}{\partial\tau}=
    \{\Pi_{\alpha},H_T \},~~~~~~\tau\equiv\sigma^0,
\end{equation}
where the Poisson bracket is defined by
\begin{equation}\label{pois}
    \{\cdot,\cdot\}:= \int d^p\sigma\Big(\frac{\partial\cdot}{\partial X^{\alpha}}
    \frac{\partial\cdot}{\partial \Pi_{\alpha}}
     - \frac{\partial\cdot}{\partial \Pi_{\alpha}}
    \frac{\partial\cdot}{\partial X^{\alpha}}\Big).
\end{equation}

One finds that the constraints satisfy the following algebra:
\begin{eqnarray}\nonumber
  \{C(f), C(g)\}= 4\mu_p^2C_i(hh^{ij}(fg_{,j}-gf_{'j})) \\
  \{C_i(f^i), C(g)\} = C(f^ig_{,i}-gf^i_{,i})\\ \nonumber
  \{C_i(f^i), C_i(g^i)\} = C_i(f^jg^i_{,j}-g^jf^i_{,j})
\end{eqnarray}
where $h_{ab}:=det[\partial_a X^{\alpha} \partial_b
    X^{\beta} g_{\alpha\beta}]$, $h:=det[h_{ab}]$ and the smeared phase space function $A(f)$ is defined
    as:
\begin{equation}
    {A}(f):= \int_{\Sigma} d^p\sigma \;
    f(\sigma^a)A(X^{\mu}, \Pi_{\mu}).
\end{equation}

\smallskip
\goodbreak

\goodbreak
\section{A $p$-brane in compactified Milne Universe}

In this section we will specialize the general formulas gathered
in previous sections to the cases of the lowest dimensional
objects, i.e. particle, string and membrane, propagating in the
compactified Milne space, $\Milne_C$. We will solve the equations
of motion in case of particle and string. We will also introduce
dimensionally reduced states that are possible for string and
membrane. These reductions will play a role in canonical
formulation, prior to quantization performed in the next chapter.

\bigskip
\goodbreak

\subsection{Particle}
For the sake of clarity we restrict the following analysis to the
significant dimensions of the $\Milne_C$ space, i.e. the time and
the disappearing/appearing dimensions. In other words, we use the
metric
\begin{equation}\label{rm}
 ds^2 =  - dt^2 +t^2 d\theta^2.
\end{equation}

\subsubsection{The Lagrangian formalism}

The Polyakov action, $S_P$, describing a relativistic test
particle of mass $m$ in a gravitational field
$g_{\alpha\beta}~~(\alpha,\beta =0,1)$ is defined by (see
(\ref{e:polyakov}) and \cite{MP6,MP8}):
\begin{equation}\label{e:polyakov:0} S_P=\int
d\tau\: L(\tau),~~~~~~~
L(\tau):=\frac{m}{2}\:(\frac{\dot{X}^{\alpha} \dot{X}^{\beta}}{e}
g_{\alpha\beta}-e),~~~~\dot{X}^{\alpha} :=dX^{\alpha}/d\tau,
\end{equation}
where $\tau$ is an evolution parameter, $e(\tau)$ denotes the
`einbein' on the world-line ($e(\tau)\equiv\sqrt{\gamma}$ in
(\ref{e:polyakov})), $X^0$ and $X^1$ are time and space
coordinates, respectively.

In the specified metric (\ref{rm}) the Lagrangian in
(\ref{e:polyakov:0}) reads
\begin{equation}\label{lag}
    L(\tau)=  \frac{m}{2e}\:(T^2 \dot{\Theta}^2 -\dot{T}^2 -e^2).
\end{equation}
For the Lagrangian (\ref{lag}) the equations of motion read
\begin{equation}\label{ruch} \frac{d}{d\tau}\bigg(\frac{m
T^2\dot{\Theta}}{e}\bigg)=0,~~~~~~
\ddot{T}-\bigg(\frac{\dot{e}}{e}\bigg)\dot{T}+\dot{\Theta}^2T=0,~~~~~~
e^2=\dot{T}^2-T^2\dot{\Theta}^2 .
\end{equation}
The solution to (\ref{ruch}) may be expressed in a gauge-invariant
manner:
\begin{equation}\label{sol} \Theta(T)=
-\int\frac{d(\frac{c_{1}}{mT})} {\sqrt{1+(\frac{c_{1}}{mT})^2}}=
-\textrm{arsinh} \bigg(\frac{c_{1}}{mT}\bigg)+c_2,~~~~c_1\in\dR
,~~0 \leq c_2 <2\pi.
\end{equation}
Now one observes that for $c_1\neq 0$ particle winds infinitely
many times around $\theta$-dimension as $t\rightarrow 0~$ and the
value of $\frac{d\Theta}{dT}$ is not well-defined for $t=0$. If we
distinguish between points of different value of $\theta$ for
$t=0$, then the particle becomes topologically (of length equal to
zero) a string at the singularity, since every point in the line
$(t,\theta)=(0,\mathbb{S}^1)$ is the $t\rightarrow 0~$ limit of
the formula (\ref{sol}). Therefore, the dynamics has no unique
extension beyond the singularity no matter which topology one
ascribes to the point(s) $t=0$.\\

We now see that there are \textit{two different aspects of
non-uniqueness} of the particle's classical propagation across the
singularity:
\begin{enumerate}
    \item There is no coordinate system covering a neighborhood
    of the singularity unless we assign the topology of circle to
    it.

    \item Even if we do this the particle cannot be traced down to
    the very singularity since it winds infinitely many times
    around the compact dimension.
\end{enumerate}

\textit{Taking into account the above one may say that only the
$c_1=0$ states can be uniquely extended beyond the singularity.}

\bigskip
\subsubsection{The Hamiltonian formalism}

In the Hamiltonian formalism we obtain the constraint (see
(\ref{conC}) and \cite{MP5}):
\begin{equation}\label{conP}
    C :=  \Pi_a\Pi_bg^{ab}+m^2=(\Pi_\theta/T)^2 - (\Pi_t)^2 + m^2,
\end{equation}
where  $\Pi_t := \partial L/\partial\dot{T}\:$ and $\Pi_\theta
:=\partial L/\partial\dot{\Theta}\:$ are canonical momenta. The
Hamiltonian $H_T=A~C$ (where $A$ is an arbitrary function of
$\tau$) gives the equations of motion:
\begin{eqnarray}
\dot{\Theta} = \frac{2A(\tau)}{T^2}\Pi_{\theta},&~~~~\dot{T}=2A(\tau)\Pi_t, \\
\dot{\Pi}_{\theta}=0,&~~~~\dot{\Pi}_t=
\frac{2A(\tau)}{T^3}\Pi_{\theta}^2.
\end{eqnarray}

\bigskip

 Thus, during  evolution of the system
$\Pi_\theta$ is conserved. Owing to the constraint (\ref{conP}),
$\Pi_t$ blows up as $T\rightarrow 0~$ for $\Pi_{\theta} \neq 0 $.
This is a real problem, i.e.  it cannot be avoided by a suitable
choice of coordinates. It is called the 'blue-shift' effect.

However, trajectories of a test particle, i.e. nonphysical
particle, coincide (by definition) with time-like geodesics of an
empty spacetime, and there is no obstacle for such geodesics to
reach/leave the singularity. It is clear that such an extension
cannot be unique because at $t=0$ the Cauchy problem for the
geodesic equation is not well defined. \textit{Therefore the
$\Pi_{\theta}=0$ states are distinguished as the only
deterministically extendable ones}.

We postpone further discussion to the next chapter, where we will
deal with quantum theory.

\subsection{String}
\subsubsection{The Lagrangian formalism}

One can check that using the embedding functions $T$ and $\Theta$
for expressing dynamics of a string even in the most convenient
gauges produces a difficult system of coupled non-linear
equations. Therefore we will proceed in a different way \cite{MP1}
and use the local flatness of the $\Milne_C$ space, a fact, that
is transparent in the coordinates:
\begin{equation}\label{min}
    x^0 = t\cosh{\theta},~~~~~~x^1 = t\sinh{\theta} .
\end{equation}
This strategy is to be effective because the solutions to the
dynamics of string in Minkowski spacetime are already known.

An action describing a test string in a fixed background spacetime
with metric $g_{\mu \nu}$ may be given by  the Polyakov action
(see (\ref{e:polyakov})):
\begin{equation}\label{ac2}
S_{P} = -\frac{1}{2}\mu_1\int d\tau d\sigma
~\sqrt{-\gamma}~\gamma^{ab}~X_{,a}^\mu X^\nu_{,b}~g_{\mu\nu} ,
\end{equation}
where $\mu_1$ is a mass per unit length, $\gamma_{ab}$ is the
string world-sheet metric, $\gamma := det[\gamma_{ab}]$ and where
$X^{\mu}=(T,X^1,\dots,X^d)$.

Inserting $\sqrt{-\gamma}~\gamma^{ab}:=\eta^{ab}$ (which is a
special choice of gauge on the string's world-sheet)  and
$g_{\mu\nu}:=\eta_{\mu\nu}$ into (\ref{ac2}) leads to, after
applying variational principle, the following equations of motion
\begin{equation}\label{eom}
\partial_{\tau}^2X^{\mu} - \partial_{\sigma}^2X^{\mu} =0 ,
\end{equation}
plus a boundary condition. Hence, the string's propagation in
Minkowski space is described by
\begin{equation}\label{mink}
  X^{\mu}(\tau,\sigma) = X^{\mu}_+(\tau+\sigma)+ X^{\mu}_-(\tau-\sigma),
 \end{equation}
\begin{equation}\label{gauge2}
  \partial_{\tau}X^{\mu}\partial_{\tau}X_{\mu}
  +\partial_{\sigma}X^{\mu}\partial_{\sigma}X_{\mu} =0,~~~~~~
\partial_{\tau}X^{\mu}\partial_{\sigma}X_{\mu}=0 ,
\end{equation}
where $X_\pm^\mu$ are any functions. The equations (\ref{gauge2})
are gauge constraints. We can make use of these solutions to
construct string solutions in the $\Milne_C$ space which wind
round the compact dimension, and therefore can be expressed in
terms of a function $\overline{X}(t,\theta)$, where
$\overline{X}:=(X^2, X^3, \texttt{\ldots}, X^d)$.

\begin{figure}[h]\label{singularmap}
\hspace{0.08\textwidth}
\begin{minipage}[b]{0.55\textwidth}
\flushleft
\includegraphics[width=0.56\textwidth,height=0.56\textwidth]{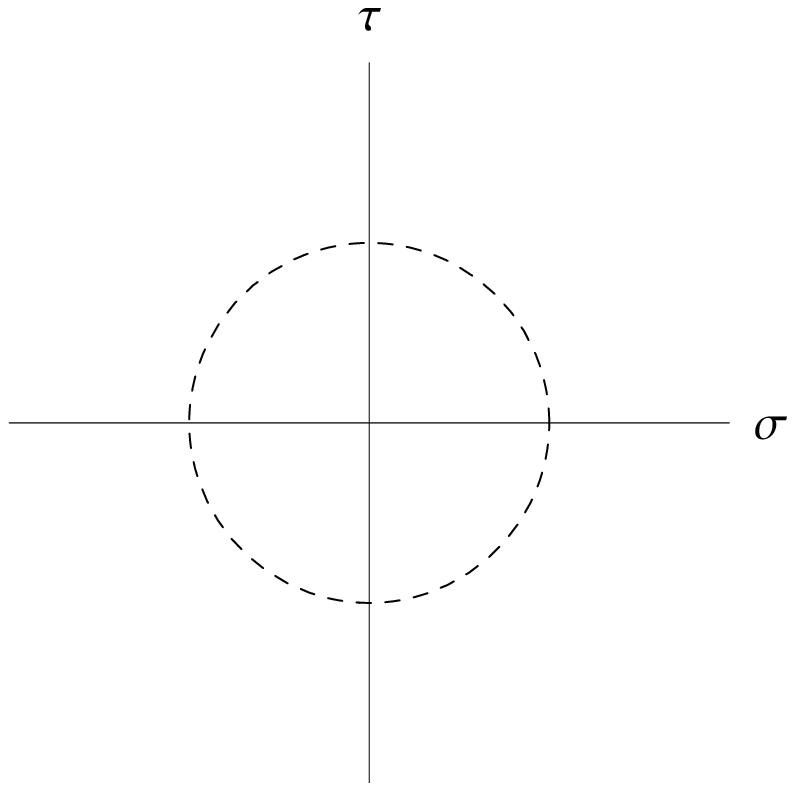}
\nolinebreak
\begin{minipage}[b]{0.32\textwidth}
\includegraphics[width=\textwidth]{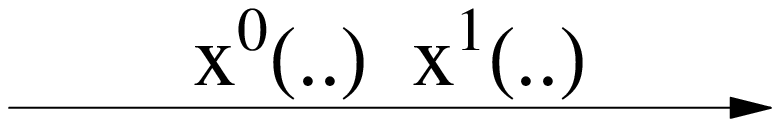}
\vspace{0.6\textwidth}
\end{minipage}
\nolinebreak
\includegraphics[width=0.56\textwidth,height=0.56\textwidth]{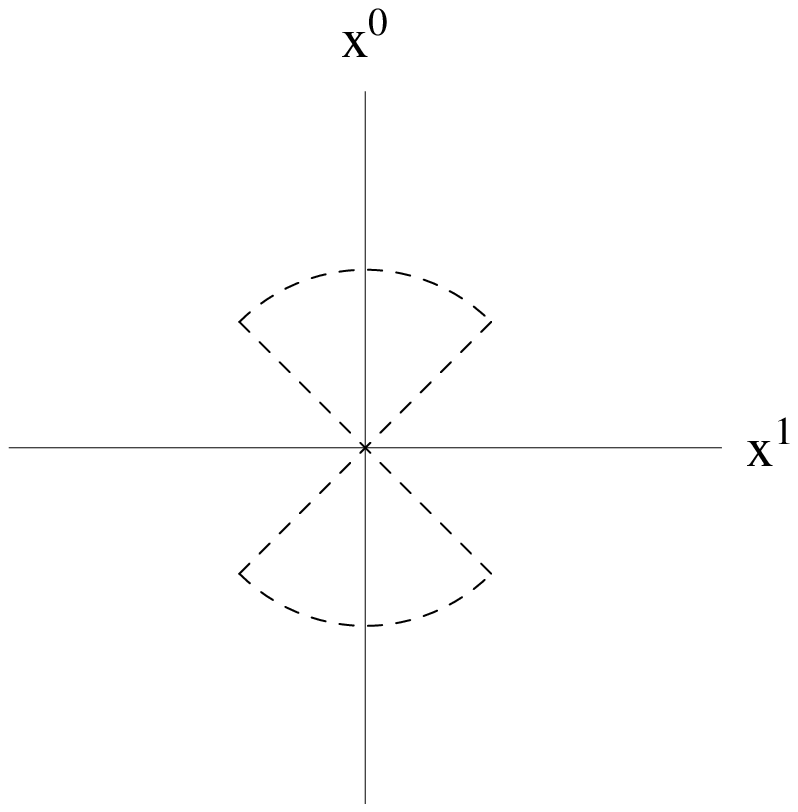}
\end{minipage}
\caption{Singular property of the map $(\tau,
\sigma)\longrightarrow ( x^0, x^1)$. The map is invertible for
$\tau \neq 0$, and non-invertible for $\tau=0$.}
\end{figure}

It follows from (\ref{min}) that the range of this mapping  has a
nontrivial topology due to the existence of the singular point
$(x^0,x^1)=(0,0)$ (see figure \ref{singularmap}). Combining this
property with the general solution (\ref{mink}), we inevitably
arrive to the following topology condition
\begin{equation}\label{sym}
  x^0 = f(\tau+\sigma)-f(-\tau+\sigma),~~~~~
  x^1 = g(\tau+\sigma)-g(-\tau+\sigma) ,
\end{equation}
where $f$ and $g$ are any functions. One can always arrive to the
above form by performing an appropriate conformal transformation
$\sigma_{\pm}\rightarrow\widetilde{\sigma_{\pm}}(\sigma_{\pm})$,
where $\sigma_{\pm}=\sigma\pm\tau$. More precisely, let us make
the conformal transformation on the solution (\ref{mink}) to get
$X^0 = f(\tau+\sigma)-f(-\tau+\sigma)$. One can verify that other
forms of $X^0$ are excluded. It follows  from (\ref{min}) that we
have the implication: $(X^0=0)\Rightarrow (X^1=0)$. This means
that for $\tau=0$ we have $X^1=0$, which leads to $X^1 =
g(\tau+\sigma)-g(-\tau+\sigma)$.

Now, let us impose the symmetry condition  on the remaining $X^k,
(k>1)$ embedding functions. Due to the assumption made earlier,
$X^k$ are functions of $t$ and $\theta$, i.e.
$X^k(\tau,\sigma)=\widetilde{X}^k(t(\tau,\sigma),\theta(\tau,\sigma))$
and are to be periodic in $\theta$. It follows from (\ref{sym})
that
\begin{equation}\label{fun1}
  \theta = \textrm{arctanh}\Big(\frac{g(\sigma_+)-g(-\sigma_-)}
  {f(\sigma_+)-f(-\sigma_-)}\Big)
\end{equation}

\begin{equation}\label{fun2}
t=\textrm{sgn}(\tau)~\sqrt{(f(\sigma_+)-f(-\sigma_-))^2-(g(\sigma_+)
-g(-\sigma_-))^2}
\end{equation}
So the symmetry condition states that
$X^k=X^{k}_+(\sigma_+)+X^{k}_-(\sigma_-)$ is periodic in
$\theta=\textrm{arctanh}(\frac{g(\sigma_+)-g(-\sigma_-)}{f(\sigma_+)-f(-\sigma_-)})$.
In other words, we should determine $X^{k}_+$ and $X^{k}_-$ from
\begin{equation}\label{deter}
  X^{k}_+(\sigma_+)+X^{k}_-(\sigma_-) = \sum_n a_n^k(t)\exp\big
(\imath\frac{2\pi
  n}{\beta}\theta\big),
\end{equation}
where $a_n^k$ are functions of $t$ whose exact form we will
discover below.  It may seem to be impossible to satisfy these
conditions. One obstacle is due to the fact that on the left-hand
side we have a sum of functions of a single variable, while on the
right-hand side there is a sum of functions which depend in a
rather complicated way on both variables. However, we can compare
both sides of (\ref{deter}) at a line. In this way one can rule
out one of the variables and compare functions dependent on just a
single variable. The procedure rests upon the fact that the
dynamics is governed by a second order differential equation
(\ref{eom}), and thus it is sufficient to satisfy the symmetry
condition by specifying $X^k$, $\partial_t X^k$ on a single
Cauchy's line. We choose it to be the singularity, i.e. the line
$\sigma_+ = -\sigma_-$, or equivalently $t=0$. One can check that
as $\sigma_++\sigma_-\rightarrow 0$, one gets $\theta \rightarrow
\textrm{arctanh} (g'/f')$, where the prime indicates
differentiation with respect to an arbitrary parameter.

Our strategy consists in the imposition of the two conditions:
\begin{eqnarray}\label{one}
\lim_{\sigma_++\sigma_-\rightarrow
0}X^k=X^{k}_+(\sigma)+X^{k}_-(\sigma) = \sum_{n}
a_n^k(0)\exp\big(\imath\frac{2\pi n} {\beta}\textrm{arctanh}\big
(\frac{g'}{f'}\big )(\sigma)\big ),
\end{eqnarray}
\begin{eqnarray}\label{two}
\lim_{\sigma_++\sigma_-\rightarrow
0}\partial_tX^k=\partial_tX^{k}_+(\sigma)+\partial_tX^{k}_-(\sigma)
= \sum_n \dot{a}_n^k(0) \exp\big(\imath\frac{2\pi
n}{\beta}\textrm{arctanh}\big (\frac{g'}{f'}\big )(\sigma)
\big).~~~~
\end{eqnarray}

In this way we get the following simplifications: (i) as we
compare functions on a line we in fact compare functions of a
single variable, (ii) since we choose the line $t=0$, we obtain a
rather simple form on the right-hand side in the form of a
periodic function of $\theta=\arctan(g'/f')$. The only remaining
work to be done is to find the operator $\partial_t$ in the limit
$~\sigma_++\sigma_-\rightarrow 0$.

One can check (see the paper \cite{MP1}) that
\begin{eqnarray}
\nonumber
  \partial_t &=& \frac{\partial_-\theta}{\partial_+t\partial_-\theta-\partial_-t\partial_+
  \theta}~\partial_+ -
  \frac{\partial_+\theta}{\partial_+t\partial_-\theta-\partial_-t\partial_+\theta}
  ~\partial_-
\end{eqnarray}
in the limit $~\sigma_++\sigma_-\rightarrow 0$ turns to be
\begin{eqnarray}
  \partial_t&\longrightarrow& \frac{1}{2\sqrt{(f')^2-(g')^2}}
  (\partial_++\partial_-)~~\Big |_{\sigma_+=-\sigma_-}.
\end{eqnarray}

Now it is straightforward to check that application the conditions
(\ref{one}) and (\ref{two}) render
\begin{eqnarray}\label{solution}
  X^0 &=& q\sinh(\sigma_+)+q\sinh(\sigma_-),\\
  X^1 &=& q\cosh(\sigma_+)-q\cosh(\sigma_-),\\
  \nonumber\\
  X^k &=& \sum_n a_{n+}^k \exp\big (\imath\frac{2\pi
  n}{\beta}\sigma_+\big )\nonumber\\ &+&\sum_n a_{n-}^k \exp\big (\imath\frac{2\pi
  n}{\beta}\sigma_-\big
  )+{c_0^k}(\sigma_++\sigma_-),
\end{eqnarray}
where $k>1$. These solutions should satisfy the gauge conditions
(\ref{gauge2}), which now takes the form
\begin{equation}\label{gauge}
  \partial_+X_k\partial_+X^k = q^2 =
  \partial_-X_k\partial_-X^k .
\end{equation}

Now one can find that the solutions as functions of $t$ and
$\theta$ have the form
\begin{eqnarray}\label{gen}
  X^k(t,\theta) &=& \sum_n \Big ( a_{n+}^k e^{\imath\frac{2\pi
  n}{\beta}\textrm{arcsinh}\big (\frac{t}{2q}\big )}+a_{n-}^k e^{-\imath\frac{2\pi
  n}{\beta}\textrm{arcsinh}\big (\frac{t}{2q}\big )}\Big )  \exp\big (\imath\frac{2\pi
  n}{\beta}\theta\big )
  \nonumber\\&+&2c_0^k\textrm{arcsinh}\Big (\frac{t}{2q}\Big ),
\end{eqnarray}
where $n$ denotes $n$-th excitation. The number of arbitrary
constants  in (\ref{gen}) can be reduced by the imposition of the
gauge condition (\ref{gauge}).

Equation (\ref{gen}) defines the solution  corresponding to the
compactification of one space dimension to $S^1$. The solution
corresponding to the compactification to a {\it segment}, can be
obtained from (\ref{gen}) by the imposition of the condition
$X^k(t,\theta) = X^k(t,-\theta)$, which leads to $a^k_n = -b^k_n$
and $\varphi_n^k = - \phi_n^k$, where $\theta \in [0,\beta/2]$.

\emph{The general solution (\ref{gen}) shows that the propagation
of a string through the cosmological singularity is not only
continuous and unique, but also analytic. Solution in the
$\Milne_C$ space is as regular as in the case of the Minkowski
space}.

The imposition of the gauge constraint (\ref{gauge}) on the
infinite set of functions given by (\ref{gen}) produces an
infinite variety of physical states. This procedure goes exactly
in the same way as for a closed string in Minkowski spacetime, but
with a smaller number of degrees of freedom due to the condition
that the string is winding around the compact dimension.

\subsubsection{The Hamiltonian formalism}

There is no need to repeat all the results from the Lagrangian
formalism in the Hamiltonian formalism. Our need for the
Hamiltonian formalism comes from our intention to quantize the
system canonically. Although we have found all the solutions for a
string winding round the compact dimension, we are going to
quantize only special states, i.e. strings which are winding
\emph{uniformly}. In this way we reduce a field theory (with
infinitely many degrees of freedom) to a mechanical system and
thus multiply our chances for success.

We analyze the dynamics of a string  in the {\it zero-mode} (the
lowest energy state) which is {\it winding} around the
$\theta$-dimension \cite{MP3,MP4}. The string in such a state is
defined by the condition
\begin{equation}\label{con1}
    \sigma^1 := \theta~~~~~~\mbox{and}~~~~~
    \partial_\theta X^\mu =0=\partial_\theta
    \Pi_\mu,
\end{equation}

One can show that the condition (\ref{con1}) eliminates the
canonical pair $(\Theta,\Pi_\theta)$ and thus reduces the
constraints (see (\ref{conC}), (\ref{conCi}) and \cite{MP3,MP4}):
\begin{equation}\label{cond1}
    C = \Pi_\mu(\tau)\;\Pi_\nu (\tau)\;\eta^{\mu\nu}
    + \check{\mu}_1^2 \;T^2(\tau)\approx 0,~~~~~~C_1 \equiv 0,
\end{equation}
where $\check{\mu}_1 \equiv \theta_0 \mu_1$ and $X^{\mu}$ no
longer includes the embedding functions corresponding to the
compact dimension $\theta$.

Let us solve the dynamics. The equations of motion (\ref{hameq})
read
\begin{equation}\label{eqp}
    \dot{\Pi}_t(\tau)= - 2 A(\tau)\;\check{\mu}_1^2\;T(\tau),~~~~~~
    \dot{\Pi}_k (\tau)= 0,
\end{equation}
and
\begin{equation}\label{eqx}
    \dot{T} (\tau)= - 2 A(\tau)\;\Pi_t(\tau),~~~~~~\dot{X}^k (\tau)=
    2 A(\tau)\;\Pi_k(\tau),
\end{equation}
where $A=A(\tau)$ is any regular function.

It can be verified that in the gauge $A(\tau)=1$, the solutions
are
\begin{equation}\label{solp}
 \Pi_t(\tau)= b_1 \exp (2\check{\mu}_1 \tau) + b_2 \exp (-2\check{\mu}_1
    \tau),~~~~~~\Pi_k(\tau)= \Pi_{0k},
\end{equation}
where $~\;b_1, b_2, \Pi_{0k} \in \dR$, and
\begin{equation}\label{solx}
    T(\tau)= a_1 \exp (2\check{\mu}_1 \tau) + a_2 \exp (-2\check{\mu}_1
    \tau),
    ~~~~~X^k (\tau)= X^k_0 + 2 \Pi_{0k} \;\tau,
\end{equation}
where $\;a_1, a_2, X^k_0 \in \dR$.

To analyze  the propagation of a string across the singularity
$t=0$, we eliminate $\tau$ from (\ref{solp}) and  (\ref{solx}).
Making the choice of $a_1$ and $a_2$ in such a way that $a_1 a_2
<0$ leads to one-to-one relation between $T$ and $\tau$. For
instance, one may  put
\begin{equation}\label{coef}
    a_1 = - a_2 = \sqrt{\Pi_0^k \Pi_{0k}}/2 \check{\mu}_1,
\end{equation}
that gives
\begin{equation}\label{elim}
      T(\tau) =  \sqrt{\Pi_0^k
     \Pi_{0k}}\;\sinh(2\check{\mu}_1\;\tau)/\check{\mu}_1 ,
\end{equation}
which can be rewritten as
\begin{equation}\label{jaw}
    \tau = \frac{1}{2 \check{\mu}_1}\sinh^{-1}\Big( \frac{\check{\mu}_1}
    {\sqrt{\Pi_0^k \Pi_{0k}}}\;t \Big),
\end{equation}
due to $T = t$. The insertion of (\ref{jaw}) into (\ref{solx})
gives
\begin{equation}\label{sols}
 X^k (t)= X^k_0 + \frac{\Pi_0^k}{ \check{\mu}_1}\sinh^{-1}\Big( \frac{\check{\mu}_1}
    {\sqrt{\Pi_0^k \Pi_{0k}}}\;t \Big).
\end{equation}
The solution (\ref{sols}) is bounded and continuous at the
singularity. Thus, the classical dynamics of the zero-mode winding
string is well defined in the $\Milne_C$ space. In fact, it
corresponds to the solution (\ref{gen}) for
\begin{equation}
q=\frac{\sqrt{\Pi_0^k
\Pi_{0k}}}{2\check{\mu}_1},~~c^k_0=\frac{\Pi^k_0}{2\check{\mu}_1},~~a^k_{0+}+a^k_{0-}=X^k_0.
\end{equation}

\textit{One may note that in case the string is winded uniformly
around the compact dimension the propagation is unique and smooth
through the singularity no matter whether it has circle or point
topology.}

Let us once more distinguish between two different
topologies one may assign to compactified Milne space, for which
the line element reads (we omit higher dimensions):
\begin{equation}
ds^2 = -dt^2  +  t^2 d\theta^2.
\end{equation}
As it is clear from the above formula, for $t=0$ the metric is
degenerate and there is a singularity. We say the singularity has
point topology if there is a single point with $t=0$. On the
contrary, we say the singularity has circle topology if there is a
continuum of points with $t=0$, each enumerated with different
value of $\theta$. Clearly, the distance between these points is
equal to zero.

\subsection{Membrane}

The case of a membrane constitutes the most difficult part of our classical analysis. We restrict ourselves to considering the states that are
winding uniformly round the compact dimension both in the Lagrangian and Hamiltonian formulation. In the Lagrangian formalism we will limit our
discussion to stating the equations of motion and laws of conservation in the gauge, which as we believe, is the most promising for finding the
solutions. In the Hamiltonian formalism the mentioned reduction leads to an algebra of two constraints, which we will rephrase in a form
convenient for Dirac's quantization.

\subsubsection{The Lagrangian formalism}

The Nambu-Goto action for a membrane in the $\Milne_C$ space reads
\begin{eqnarray}\nonumber
    S_{NG}&=&-\mu_2\int d^3\sigma\sqrt{-det(\partial_aX^{\mu}\partial_bX^{\nu}
    g_{\mu\nu})}\\ \label{membrane}
    &=&-\mu_2\int
d^3\sigma\sqrt{-det(-\partial_aT\partial_bT+T^2\partial_a\Theta\partial_b
\Theta+\partial_aX^k\partial_bX_k )}
\end{eqnarray}
where $(T,\Theta ,X^k)$ are embedding functions of the membrane
corresponding to the spacetime coordinates $(t,\theta,x^k)$
respectively.

An action  $S_{NG}$ in the lowest energy winding mode, defined by
(\ref{con1}), has the form \cite{MP9}
\begin{eqnarray}\nonumber
    S_{NG}&=&-\mu_2\theta_0\int
d^2\sigma\sqrt{-T^2det(-\partial_aT\partial_bT+\partial_aX^k\partial_bX_k
)}\\ \label{string2} &=&-\mu_2\theta_0\int
d^2\sigma\sqrt{-det(\partial_aX^{\alpha}\partial_bX^{\beta}\widetilde{g}_{\alpha\beta})}.
\end{eqnarray}
where $a,b\in \{0,1\}$,\;
$\widetilde{g}_{\alpha\beta}=T\eta_{\alpha\beta}$ and
$\theta_0=\int d\theta$. Now it is clear that the dynamics of a
{\it membrane} in the state (\ref{con1}) is equivalent to the
dynamics of a {\it string} with tension $\mu_2\theta_0$ in the
spacetime with the metric $\widetilde{g}_{\alpha\beta}$.

The Nambu-Goto action (\ref{string2}) is equivalent to the
Polyakov action

\begin{equation}\label{Pstring}
    S_p=-\frac{1}{2}\mu_2\theta_0\int d^2\sigma\sqrt{\gamma}(\gamma^{ab}
    \partial_aX^{\alpha}\partial_bX^{\beta}~T\eta_{\alpha\beta})
\end{equation}
because variation with respect to $\gamma^{ab}$ (and using
$\delta\gamma=\gamma\gamma^{ab}\delta\gamma_{ab}$) gives
\begin{equation}\label{gamma}
    \partial_aX^{\alpha}\partial_bX^{\beta}~T\eta_{\alpha\beta}-\frac{1}{2}\gamma_{ab}
    \gamma^{cd}\partial_cX^{\alpha}\partial_dX^{\beta}~T\eta_{\alpha\beta}=0.
\end{equation}
The insertion of (\ref{gamma}) into the Polyakov action
(\ref{Pstring}) reproduces the Nambu-Goto action (\ref{string2}).

In the gauge $\sqrt{-\gamma}\gamma^{ab}=1-\delta_{ab}$ the action
(\ref{Pstring}) reads
\begin{equation}\label{Pstring2}
    S_p=-\mu_2\theta_0\int d^2\sigma(\partial_+X^{\alpha}\partial_-
    X^{\beta}~T\eta_{\alpha\beta})
\end{equation}
where $\partial_{\pm}=\frac{\partial}{\partial {\sigma_{\pm}}}$.

The least action principle applied to (\ref{Pstring2}) gives the
following equations of motion
\begin{eqnarray}\label{Leom1}
    \partial_{-}(T\partial_{+}X^k)+\partial_{+}(T\partial_{-}X^k)=0\\ \label{Leom2}
\partial_{-}(T\partial_{+}T)+\partial_{+}(T\partial_{-}T)+\partial_+X^{\alpha}
\partial_-X^{\beta}~\eta_{\alpha\beta}=0,
\end{eqnarray}
where (\ref{gamma}) with the specified gauge reads
\begin{equation}\label{constraint}
    \partial_+X^{\alpha}\partial_+X^{\beta}~\eta_{\alpha\beta} = 0 = \partial_-
    X^{\alpha}\partial_-X^{\beta}~\eta_{\alpha\beta}.
\end{equation}

On the other hand, the action (\ref{Pstring2}) is invariant under
the conformal transformations, i.e.
$\sigma_{\pm}\longrightarrow\sigma_{\pm}+{\epsilon}_{\pm}(\sigma_{\pm})$.
It is so because for such transformations we have $\delta
X^{\alpha}=-{\epsilon}_{-}\partial_-X^{\alpha}-{\epsilon}_{+}\partial_+X^{\alpha}$
and hence
\begin{equation}\label{K}
    \delta S_p=-\mu_2\theta_0\int d^2\sigma \bigg(
    \partial_-(-{\epsilon}_{-}\partial_+X^{\alpha}\partial_-X^{\beta}~
    T\eta_{\alpha\beta})+\partial_+(-{\epsilon}_{+}\partial_+X^{\alpha}
    \partial_-X^{\beta}~T\eta_{\alpha\beta})\bigg),
\end{equation}
which is equal to zero since the fields $X^{\alpha}$ either vanish
at infinity or are periodic. Now let assume that the fields
$X^{\alpha}$ satisfy (\ref{Leom1}) and (\ref{Leom2}). Then
(\ref{K}) can be rewritten as
\begin{eqnarray}\nonumber
    \delta S_p&=&-\mu_2\theta_0\int d^2\sigma \big(
    \partial_-(-{\epsilon}_{-}\partial_+X^{\alpha}\partial_-X^{\beta}~
    T\eta_{\alpha\beta})+\partial_+(-{\epsilon}_{-}\partial_-X^{\alpha}
    \partial_-X^{\beta}~T\eta_{\alpha\beta})\\
\label{minimal}&+&~\partial_+(-{\epsilon}_{+}\partial_+X^{\alpha}\partial_-
X^{\beta}~T\eta_{\alpha\beta})+\partial_-(-{\epsilon}_{+}\partial_+X^{\alpha}
\partial_+X^{\beta}~T\eta_{\alpha\beta})\big)
\end{eqnarray}
which leads to the conservation of currents
\begin{equation}\label{noether}
  \partial_-T_{++}=0,~~~~~~\partial_+T_{--}=0
\end{equation}
where
\begin{equation}\label{currrent}
    T_{++}={\epsilon}_{+}\partial_+X^{\alpha}\partial_+X^{\beta}~T\eta_{\alpha\beta},
    ~~~~T_{--}={\epsilon}_{-}\partial_-X^{\alpha}\partial_-X^{\beta}~T\eta_{\alpha\beta}~.
\end{equation}
One can verify that the vector fields ${\epsilon}_{-}\partial_-$
and ${\epsilon}_{+}\partial_+$  satisfy the following  Lie algebra
\begin{equation}\label{algebraN1}
[ f_+\partial_+ , g_+\partial_+
]=(f_+\acute{g}_+-g_+\acute{f}_+)\partial_+,
\end{equation}
\begin{equation}\label{algebraN2}
[ f_-\partial_- , g_-\partial_-
]=(f_-\acute{g}_--g_-\acute{f}_-)\partial_-,
\end{equation}
\begin{equation}\label{algebraN3}
[ f_+\partial_+ , g_-\partial_- ]=0.
\end{equation}

\subsubsection{The Hamiltonian formalism}

From the general theory described in the section \ref{s:ham} we
know that in the case of membrane the system is described by three
first-class constraints, which close to Poisson algebra with
structure functions on the phase space. However, \emph{little is
known about representations of algebras of such type}. Therefore
we will consider only the membranes in the lowest energy winding
mode defined in eq. (\ref{con1}) and subsequently by the action
(\ref{string2}). These reduced states are mathematically
equivalent to strings propagating in the curved spacetime with the
metric $g_{\alpha\beta}=|T|n_{\alpha\beta}$ and thus are
characterized by two constraints. The Hamiltonian corresponding to
the action (\ref{string2}) has the form:
\begin{equation}\label{hamiltonian}
H_T = \int d\sigma \mathcal{H}_T,~~~~\mathcal{H}_T := A C + A^1
C_1,
\end{equation}
where
\begin{equation}\label{con}
    C:=\frac{1}{2\mu_2\theta_0  T}\Pi_{\alpha} \Pi_{\beta}\eta^{\alpha\beta} +
    \frac{\mu_2\theta_0}{2} \;T\;\partial_a X^{\alpha} \partial_b
    X^{\beta} \eta_{\alpha\beta}\approx 0,~~~~
    C_1 := \partial_{\sigma} X^{\alpha} \Pi_{\alpha} \approx 0
\end{equation}
and  $A=A(\tau,\sigma)$ and $A^1 = A^1(\tau,\sigma)$ are any
regular functions. The constraint $C$ and $C_1$ may be interpreted
as diffeomorphism generators in the space of solutions to
Hamilton's equations (see the paper \cite{MP9}). Let us redefine
the constraints in the following way:
\begin{equation}\label{red}
C_{\pm}:=\frac{C\pm C_1}{2}
\end{equation}
and check that they close to the Lie algebra:
\begin{equation}\label{Alg1}
    \{ \check{C}_{+}(f),
    \check{C}_{+}(g)\} = \check{C}_{+}(f\acute{g}-g\acute{f}),
\end{equation}
\begin{equation}\label{Alg2}
    \{ \check{C}_{-}(f),
    \check{C}_{-}(g)\} = \check{C}_{-}(f\acute{g}-g\acute{f}),
\end{equation}
\begin{equation}\label{Alg3}
\{ \check{C}_{+}(f), \check{C}_{-}(g)\} = 0 .
\end{equation}
Let us use the functions of the form $\exp(\imath n\sigma)$ as the
basis in the space of the smearing functions $f$ and $g$, so the
above algebra gains the compact form:
\begin{equation}
[L^+_n,L^+_m] = \imath (m-n)L^+_{n+m}
\end{equation}
\begin{equation}
[L^{-}_n,L^{-}_m] =\imath (m-n)L^{-}_{n+m}
\end{equation}
\begin{equation}
[L^+_n,L^-_m]= 0
\end{equation}
where $L^{\pm}_n=\check{C}_{\pm}(e^{\imath n\sigma})$ for $n,m\in\mathbb{Z}$, and $\overline{L}^{\pm}_n=L^{\pm}_{-n}$.

%% file: T2.tex
\def\baselinestretch{1}

\chapter{Dirac quantization of dynamics of extended objects}

\def\baselinestretch{1.66}


In this chapter we will first briefly enumerate the essential steps in the Dirac method of quantization of constrained systems and then try to
apply Dirac's prescription to the problem of quantizing dynamics of particle, string and membrane in the $\Milne_C$ space.

\goodbreak
\section{Introduction}

Dirac proposed in \cite{PAM} a method for canonical quantization
of dynamics of constrained systems, according to which:
\begin{enumerate}
    \item First, one applies stabilization
algorithm, so one obtains a Hamiltonian $\mathcal{H}$ and
first-class constraints $\mathcal{O}_a,~a=0,1,\dots$, which by
definition close to Poisson algebra; sometimes reduction in number
of conjugate pairs and thus number of first-class constraints is
possible.
    \item Then, one constructs a (essentially) self-adjoint representation for the algebra of the Hamiltonian and the
observables so that their commutators resemble the standard canonical prescription i.e. $\widehat{\{A,B \}}=\frac{1}{\imath}[\hat{A},\hat{B}]$.
    \item  Subsequently, one finds the intersection
of all the kernels of the constraints, i.e. $\{~\bigcap_a
\mathcal{K}_a :~\Psi\in
\mathcal{K}_a\Leftrightarrow\hat{\mathcal{O}}_a\Psi=0~\}$.
    \item Finally one introduces a Hilbert space structure
    on the intersection $\bigcap_a
\mathcal{K}_a$, i.e. one redefines scalar product, because the kernel usually does not belong to the starting Hilbert space. One can do it e.g.
by applying so called group-averaging method \cite{GAM2,GAM1}. Below we do not modify definition of scalar product since we restrict our model
only to the neighborhood of singularity, which gives the upper limit for the length of any time-like curve and thus guarantees
square-integrability of vectors from the kernel.

\end{enumerate}
It is worth noting that in case the Hamiltonian is a sum of
first-class constraints then there may arise questions concerning
the flow of time, since the Hamiltonian has gone with the Dirac
procedure.

\goodbreak
\section{Quantum particle}
First we will construct the quantum Hamiltonian of a particle from
the classical one (\ref{conP}). We use the following mapping (see,
e.g. \cite{Ryan:2004vv})
\begin{equation}\label{mapp}
    \Pi_k \Pi_l g^{kl} \longrightarrow \Box  :=
    (-g)^{-1/2}\partial_k [(-g)^{1/2} g^{kl} \partial_l ],
\end{equation}
where $g:=det [g_{kl}]$ and $\partial_k := \partial/\partial x^k$.
The Laplace-Beltrami operator, $\Box$,  is invariant under the
change of spacetime coordinates and it leads to  Hamiltonians that
give results consistent with  experiments  \cite{Ryan:2004vv}, and
which has been used in theoretical cosmology (see,
\cite{Turok:2004gb} and references therein).

In the case of the $\Milne_C$ space the quantum Hamiltonian,  for
$t<0$ or $t>0$, reads \cite{MP5}
\begin{equation}\label{nh}
\hat{H} = \Box + m^2 =  \frac{\partial}{\partial t^2} +
      \frac{1}{t}\frac{\partial}{\partial t} - \frac{1}{t^2}\frac{\partial^2}
      {\partial \theta^2} + m^2 .
\end{equation}
The operator $\hat{H}$ was obtained by making use of (\ref{conP})
and the gauge $A(\tau)=1$\footnote{Since the theory we use is
gauge invariant, the different choice of the gauge should not
effect physical results.}. Thus the Dirac quantization scheme
\cite{PAM,MHT} leads to the equation
\begin{equation}\label{hcn}
\hat{H} \psi(\theta,t)= 0.
\end{equation}
Let us find the non-zero solutions of (\ref{hcn}). Separating the
variables
\begin{equation}\label{sep}
  \psi(\theta,t):= A(\theta)\;B(t)
\end{equation}
leads to the equations
\begin{equation}\label{eqth}
 d^2 A/d\theta^2 + \rho^2 A = 0,~~~~\rho\in
\dR
\end{equation}
and
\begin{equation}\label{eqt}
    \frac{d^2 B}{d t^2}+ \frac{1}{t} \:\frac{dB}{dt} + \frac{m^2 t^2 +\rho ^2}{t^2}\;B =
    0,~~~~t\neq 0,
\end{equation}
where $\rho$ is a constant of separation. Two independent
continuous solutions  on  $\dS^1$ read
\begin{equation}\label{solth}
     A_1(\rho,\theta)= a_1 \cos(\rho\theta),~~~~A_2(\rho,\theta)= a_2
     \sin(\rho\theta),~~~~~~~a_1, a_2 \in \dR .
\end{equation}
Two independent solutions  on $ \dR $ (for $t<0$ or $t>0$) have
the form  \cite{Arfken:2005,SWM}
\begin{equation}\label{solt}
  B_1(\rho,t)= b_1  \Re J(i\rho,mt),~~~~B_2(\rho,t)= b_2 \Re Y(i\rho,mt),~~~~~~~b_1,
  b_2 \in \dC ,
\end{equation}
where $\Re J$ and $\Re Y$ are  the real parts of Bessel's and
Neumann's functions, respectively. Since $\rho\in\dR$, the number
of independent solutions is: $2 \times 2 \times \infty$ ( for
$t<0$ and $t>0$).

We define the scalar product on the space of solutions given by
the formulas (\ref{solth}) and (\ref{solt}) as follows
\begin{equation}\label{scalar}
    <\psi_1|\psi_2> := \int_{\widetilde{\Gamma}} d \mu \;\overline{\psi}_1 \;\psi_2,~~~~~~d\mu
    :=\sqrt{-g}\; d\theta \;dt = |t|\; d\theta \;dt,
\end{equation}
where $\widetilde{\Gamma}:= [-T,0[ \times \dS^1$ (with $T
>0$) in the pre-singulaity epoch, and $\widetilde{\Gamma}:= ]0,T]
\times \dS^1$  in the post-singularity epoch. We assume that the
$\Milne_C$ space can be used to model the universe only  during
its quantum phase, which lasts the period $[-T, T$]. No boundary
conditions on a wavefunction is imposed.

Now we construct an orthonormal basis, in the left neighborhood of
the cosmological singularity,  out of the solutions (\ref{solth})
and (\ref{solt}). One can  verify that the solutions (\ref{solth})
are orthonormal and continuous on $\dS^1$ if $\;a_1  =
\sqrt{\frac{2}{\theta_0}}= a_2\;$ and $\frac{\theta_0\rho}{2\pi} =
0,\pm 1,\pm 2,\ldots$. Some effort is needed to construct the set
of orthonormal functions out of $\Re J(i\rho,mt)$ and $\Re
Y(i\rho,mt)$. First, one may verify that these functions are
square-integrable on the interval $[-T,0]$. This is due to the
choice of the measure in the scalar product (\ref{scalar}), which
leads to the boundedness of the corresponding integrants. Second,
having normalizable set of four independent functions, for each
$\rho$, we can turn it into an orthonormal set by making use of
the Gram-Schmidt procedure (see, e.g. \cite{Arfken:2005}). Our
orthonormal and countable set of functions may be used to define
the span $\mathcal{F}$. The completion of $\mathcal{F}$ in the
norm induced by the scalar product (\ref{scalar}) defines the
Hilbert spaces $L^2(\widetilde{\Gamma} \times \dS^1,d\mu)$. It is
clear that the same procedure applies to the right neighborhood of
the singularity.

\begin{figure}[h]
\centering \subfigure[\ near the singularity]{

\includegraphics[width=2.3in]{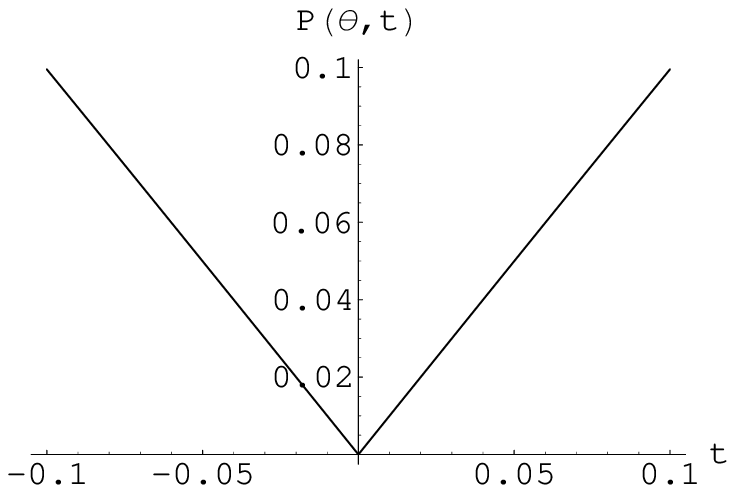}}
\hspace{0.4in} \subfigure[\ on large scale]{

\includegraphics[width=2.3in]{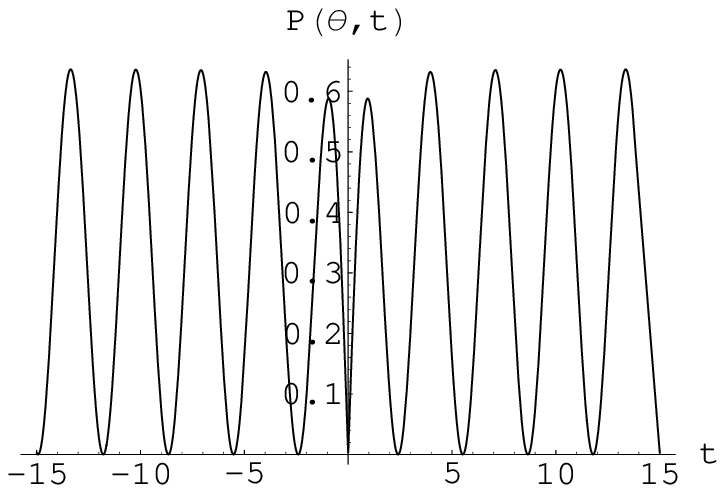}}
\caption{Probability density corresponding to $\psi(\theta,t)=
A_1(0,\theta)\;\Re J(0,t)$}

\end{figure}

\begin{figure}[h]
\centering \subfigure[\ near the singularity]{

\includegraphics[width=2.3in]{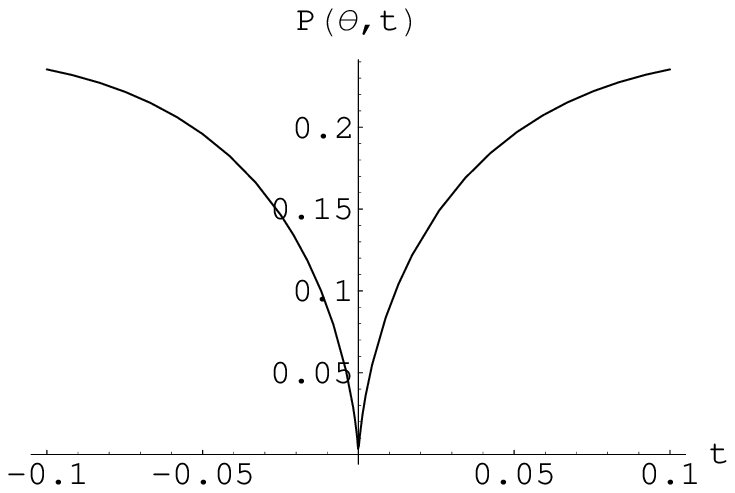}}
\hspace{0.4in} \subfigure[\ on large scale]{

\includegraphics[width=2.3in]{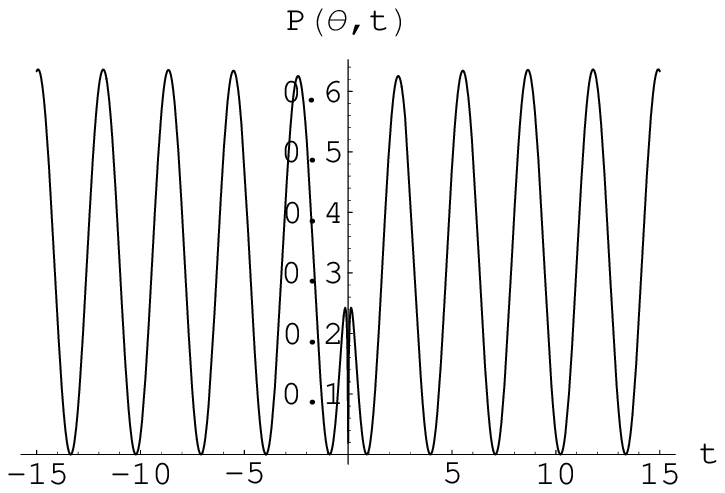}}
\caption{Probability density corresponding to $\psi(\theta,t)=
A_1(0,\theta)\;\Re Y(0,t)$}
\end{figure}

We have constructed the two  Hilbert spaces: one for the
pre-singularity epoch, $\mathcal{H}^{(-)}$, and another one to
describe the post-singularity epoch, $\mathcal{H}^{(+)}$. Next
problem is to `glue' them into a single Hilbert space,
$\mathcal{H}=L^2([-T,T] \times \dS^1,d\mu)$, that is needed to
describe the entire quantum phase. From the mathematical point of
view the gluing seems to be problematic because the Cauchy problem
for the equation (\ref{hcn}) is not well defined\footnote{Except
one case discussed later.} at $t=0$, and because we have assumed
that $t\neq 0$ in the process of separation of variables to get
Eqs. (\ref{eqth}) and (\ref{eqt}). However, arguing based on the
physics of the problem enables the gluing. First of all we have
already agreed that a \textit{classical} test particle is able to
go across the singularity (see, section 1.3.1). One can also
verify that the probability density
\begin{equation}\label{amp}
    P(\theta,t):= \sqrt{-g}\;|\psi(\theta,t)|^2 = |t|\;|\psi(\theta,t)|^2
\end{equation}
is bounded  and continuous in the domain $\;[-T,T] \times \dS^1$.
Figures 2.1 and 2.2 illustrate the behavior of $P(\theta,t)$ for
two examples of gluing the  solutions having $\rho =0$. The cases
with $\rho \neq 0$ have similar properties. Thus, the assumption
that the gluing is possible is justified. However one can glue the
two Hilbert spaces in more than one way. In what follows we
present two cases, which are radically different.

\subsubsection{Deterministic propagation}

Among all solutions (\ref{solt}) there is one, corresponding to
$\rho =0$, that attracts an attention. It reads (see e.g.
\cite{SWM})
\begin{equation}\label{n1B}
    B_1(0,mt)= b_1\;\Re J(0,mt),~~~~~~b_1\in \dR ,
\end{equation}
and has the following power series expansion close to $t=0$
\begin{equation}\label{psn}
   B_1(0,x)/b_1 = 1- \frac{x^2}{4}+\frac{x^4}{64} - \frac{x^6}{2304}
   + \mathcal{O}[x^8] .
\end{equation}
It is visualized in Fig. 2.3(a). The solution (\ref{n1B}) is smooth at the singularity, in spite of the fact that (\ref{eqt}) is singular at
$t=0$.

\begin{figure}[h]
\centering \subfigure[]{

\includegraphics[width=2.3in]{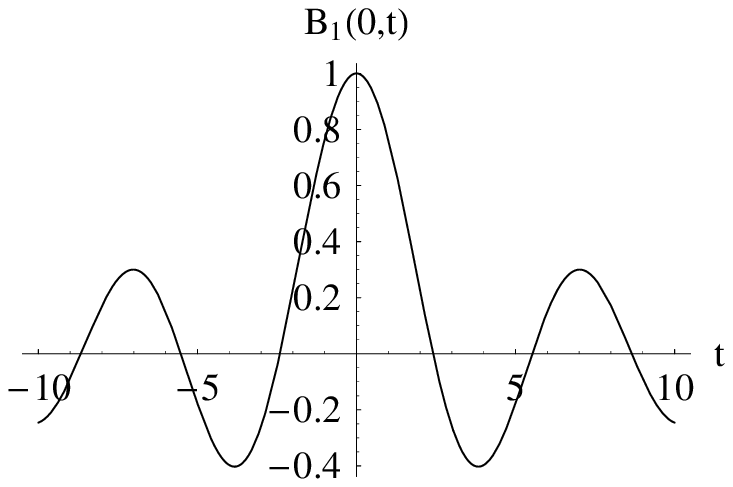}}
\hspace{0.4in} \subfigure[]{

\includegraphics[width=2.3in]{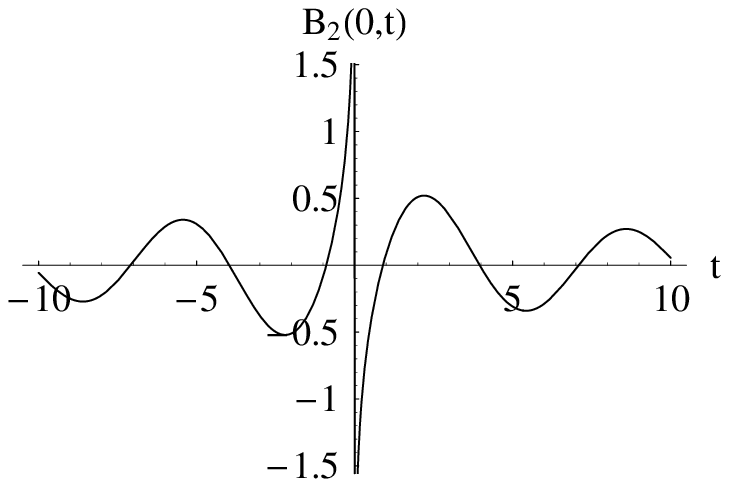}}
\caption{Continuous (a) and singular (b) propagations of a
particle with $\rho =0$. }
\end{figure}

It defines a solution to (\ref{hcn}) that does not depend on
$\theta$. Thus, it is unsensitive to the problem that one cannot
choose a common coordinate system for both $t<0$ and $t>0$.

The solution $B_1$ can be used to construct a one-dimensional
Hilbert space $\mathcal{H}=L^2([-T,T] \times \dS^1,d\mu)$. The
scalar product is defined by (\ref{scalar}) with
$\widetilde{\Gamma}$ replaced by $\Gamma := [-T,T] \times \dS^1$.

The solution (\ref{n1B}) is {\it continuous} at the singularity. It describes an unambiguous propagation of a quantum particle. Thus, we call it
the {\it deterministic} propagation.

Since (\ref{eqt}) is a second order differential equation, it
should have two independent solutions. However, the second
solution cannot be continuous at $t=0$. One may argue as follows:
The solution (\ref{n1B}) may be obtained by ignoring the
restriction $t \neq 0$ and solving (\ref{eqt}) for $\rho=0$ with
the following initial conditions
\begin{equation}\label{incon}
    B(0,0)=1,~~~~~~dB(0,0)/dt = 0.
\end{equation}
Equations (\ref{eqt}) and (\ref{incon}) are consistent, because
the middle term of the l.h.s. of (\ref{eqt}) is equal to zero due
to (\ref{incon}) so the resulting equation would be non-singular
at $t=0$. Another independent initial condition would be of the
form $\;dB(0,0)/dt \neq 0\;$. Thus, it could not lead to the
solution which is continuous at $t=0$.

\subsubsection{Indeterministic propagation}

All solutions   (\ref{solt}), except (\ref{n1B}), are
discontinuous at $t=0$. This property is connected with the
singularity of (\ref{eqt}) at $t=0$. It is clear that due to such
an obstacle the identification of corresponding solutions on both
sides of the singularity is impossible. However there  are two
natural constructions of a Hilbert space out of
$\mathcal{H}^{(-)}$ and
$\mathcal{H}^{(+)}$ which one can apply:\\
{\it (a) Tensor product of Hilbert spaces} \\
The Hilbert space is defined in a standard way \cite{EP} as
$\mathcal{H}:= \mathcal{H}^{(-)}\otimes\mathcal{H}^{(+)}$ and it
consists of functions of the form
\begin{equation}\label{tp}
    f(t_1,\theta_1;t_2,\theta_2) \equiv (f^{(-)}\otimes f^{(+)})(t_1,\theta_1;t_2,\theta_2)
    := f^{(-)}(t_1,\theta_1)\;f^{(+)}(t_2,\theta_2) ,
\end{equation}
where $f^{(-)}\in \mathcal{H}^{(-)}$ and $f^{(+)}\in
\mathcal{H}^{(+)}$. The scalar product reads
\begin{equation}\label{sten}
    <f\;|\;g>:= <f^{(-)}|\;g^{(-)}>\;<f^{(+)}|\;g^{(+)}> ,
\end{equation}
where
\begin{equation}\label{stm}
  <f^{(-)}|\;g^{(-)}>:=\int_{-T}^0 dt_1 \int_0^{2\pi}d \theta_1
  \;|t_1|\; f^{(-)}(t_1,\theta_1)\; g^{(-)}(t_1,\theta_1)
\end{equation}
and
\begin{equation}\label{stp}
  <f^{(+)}|\;g^{(+)}>:=\int_0^{T} dt_2 \int_0^{2\pi}d \theta_2
  \;|t_2|\; f^{(+)}(t_2,\theta_2)\; g^{(+)}(t_2,\theta_2) .
\end{equation}
The action of the Hamiltonian is defined by
\begin{equation}\label{mamt}
  \hat{H} \big(f^{(-)}\otimes f^{(+)}\big):= \big(\hat{H} f^{(-)}\big)
  \otimes f^{(+)} +  f^{(-)}\otimes \big(\hat{H} f^{(+)}\big).
\end{equation}

The quantum system described in this way appears to consist of two
independent parts. In fact it describes the same  quantum particle
but in two subsequent time intervals separated by the
singularity at $t=0$.\\
{\it (b) Direct sum of Hilbert spaces}\\
Another standard way \cite{EP} of defining the Hilbert space is
$\mathcal{H}:= \mathcal{H}^{(-)}\bigoplus\mathcal{H}^{(+)}$. The
scalar product reads
\begin{equation}\label{dssc}
    <f_1|f_2>:= <f_1^{(-)}|f_2^{(-)}> + <f_1^{(+)}|f_2^{(+)}> ,
\end{equation}
where
\begin{equation}\label{dsf}
    f_k := (f_k^{(-)},f_k^{(+)}) \in
    \mathcal{H}^{(-)}\times\mathcal{H}^{(+)},~~~~~~k=1,2,
\end{equation}
and where $f_k^{(-)}$ and $f_k^{(+)}$ are two completely
independent solutions in the pre-singularity and post-singularity
epochs, respectively. (The r.h.s of (\ref{dssc}) is defined by
(\ref{stm}) and (\ref{stp}).)

The Hamiltonian action on $\mathcal{H}$ reads
\begin{equation}\label{hamds}
\mathcal{H}\ni (f^{(-)},f^{(+)})\longrightarrow \hat{H}
(f^{(-)},f^{(+)}):= (\hat{H} f^{(-)},\hat{H} f^{(+)}) \in
\mathcal{H}.
\end{equation}

By the construction, the space
$\mathcal{H}^{(-)}\bigoplus\mathcal{H}^{(+)}$ includes vectors
like $(f^{(-)},0)$ and $(0,f^{(+)})$, which give non-vanishing
contribution to (\ref{dssc}) (but yield zero in case
(\ref{sten})). The former state describes the annihilation of a
particle at $t=0$. The latter corresponds to the creation of a
particle at the singularity. These type of states do not describe
the propagation of a particle {\it across} the singularity. The
annihilation/creation of a massive particle would change the
background. Such events should be eliminated from our model
because we consider a {\it test} particle which, by definition,
cannot modify the background spacetime.  Since $\mathcal{H}^{(-)}$
and $\mathcal{H}^{(+)}$, being vector spaces, must include the
zero solutions, the Hilbert space
$\mathcal{H}^{(-)}\bigoplus\mathcal{H}^{(+)}$ cannot model the
quantum phase of our system.

\goodbreak
\section{Quantum string}

In the gauge $A=1$, the Hamiltonian of a string (\ref{cond1}) is
\begin{equation}\label{ham1}
    H_T  = C = \Pi_\mu(\tau)\;\Pi_\nu (\tau)\;\eta^{\mu\nu}
    + \check{\mu}_1^2 \;T^2 .
\end{equation}
The quantum Hamiltonian corresponding to (\ref{ham1}) has the form
\cite{MP3,MP4}
\begin{equation}\label{qham1}
    \hat{H}_T = \frac{\partial^2}{\partial t^2}- \frac{\partial^2}
    {\partial x^k \partial x_k} + \check{\mu}_1^2 t^2 .
\end{equation}

According to the Dirac quantization method \cite{PAM,MHT} the
physical states $\psi$ should first of all satisfy the equation
\begin{equation}\label{eqs}
   \hat{H}_T \;\psi (t,x^k) = 0.
\end{equation}
To solve (\ref{eqs}),  we make the substitution
\begin{equation}\label{sub}
    \psi(t,x^1,\ldots,x^{d-1})= F(t)\;G_1(x^1)\;G_2(x^2)\cdots
    G_{d-1}(x^{d-1}),
\end{equation}
which turns (\ref{eqs}) into the following set of equations

\begin{equation}\label{eqG}
   \frac{d^2 G_k (q_k,x_k)}{dx_k^2}+ q_k^2 \;G_k(q_k,x_k) = 0,~~~~~~~k=1,
   \ldots,d-1 ,
\end{equation}
\begin{equation}\label{eqF}
    \frac{d^2 F(q,t)}{dt^2}+(\check{\mu}_1^2 t^2 + q^2)\;F(q,t)
    =0,~~~~~~q^2 := q_1^2 +\ldots +q_{d-1}^2,
\end{equation}
where $q_k^2, q^2 \in \dR$  are the separation constants. Two
independent  solutions to (\ref{eqG}) have the form
\begin{equation}\label{solG}
G_{1k}(q_k,x_k)= \cos (q_k x^k),~~~~~G_{2k}(q_k,x_k)= \sin (q_k
x^k),~~~~~~~~k=1,\ldots,d-1
\end{equation}
(there is no summation in $~q_k x^k~$ with respect to $k$).

Two  independent  solutions of (\ref{eqF}) read \cite{SWM}
\begin{equation}\label{solF1}
   F_1 (q,t)= \exp{(-i \check{\mu}_1 t^2/2)}\;H \Big(-\frac{\check{\mu}_1+iq^2 }
   {2 \check{\mu}_1}, (-1)^{1/4}\;\sqrt{\check{\mu}_1} \;t \Big),
\end{equation}
\begin{equation}\label{solF2}
 F_2 (q,t)=  \exp{(-i \check{\mu}_1 t^2/2)}\;_1F_1
 \Big(\frac{\check{\mu}_1+iq^2 }{4 \check{\mu}_1},\frac{1}{2},i
\check{\mu}_1 t^2 \Big),
\end{equation}
where $H(a,t)$ is the Hermite function and $_1F_1(a,b,t)$ denotes
the Kummer confluent hypergeometric function.

In what follows we present the construction of a Hilbert space,
$\mathcal{H}$,  of our system based on the solutions
(\ref{solG})-(\ref{solF2}):

\noindent First, we intend to redefine (\ref{solF1}) and
(\ref{solF2}) to get {\it bounded} functions on $\dR \times
[-t_0,t_0]$, where $[-t_0,t_0]$ denotes the `time-like'
neighborhood of the singularity.  For fixed value of $q$ and
$t\in[-t_0,t_0]$ the solutions (\ref{solF1}) and (\ref{solF2}) are
bounded functions, as it is demonstrated by the plots of  Fig. 2.4
.

\begin{figure}[h]
\centering \subfigure {\includegraphics[width=2.3in]{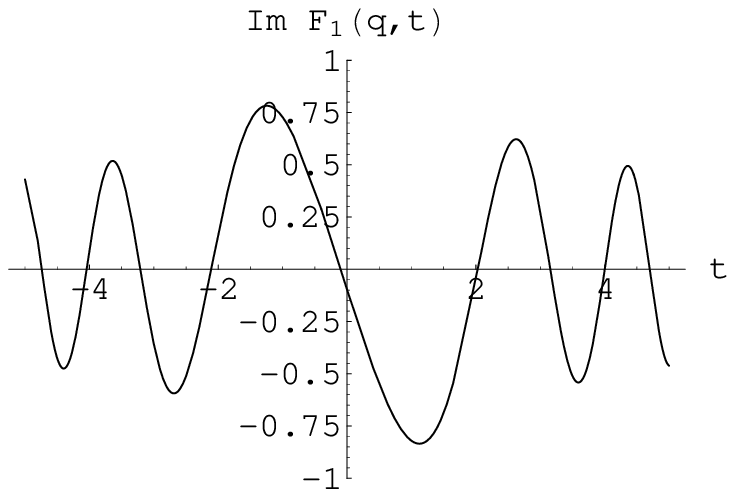}}
\hspace{0.4in} \subfigure
{\includegraphics[width=2.3in]{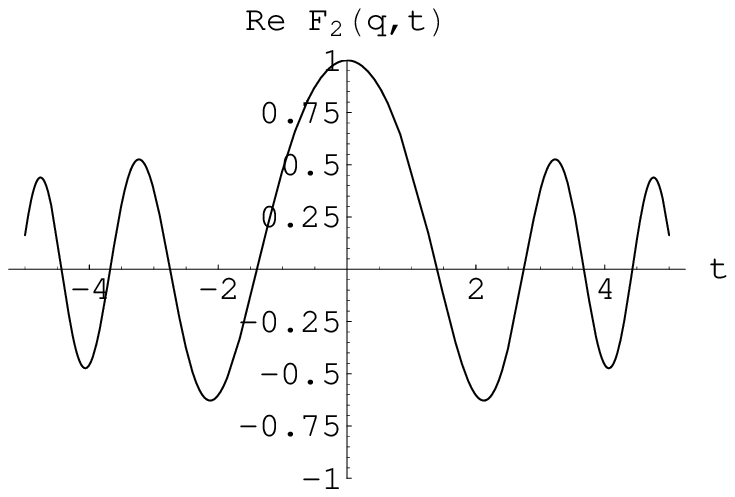}} \caption{Solutions as
functions of $t$ in the neighborhood of the singularity
($\check{\mu}_1=1,$  $q=1$).}
\end{figure}

For $q^2\gg \check{\mu}_1^2 t_0^2$, the solution to (\ref{eqF})
can be approximated by

\begin{equation}\label{solFbigq}
F(q,t)\approx A(q)\sin(qt)+B(q)\cos(qt),
\end{equation}
where $A(q)$ and $B(q)$ are any functions. Finding bounded $A(q)$
and $B(q)$ in (\ref{solFbigq}) gives bounded $F(q,t)$.  They can
be determined from  the equations ($q^2\gg \check{\mu}^2t_0^2$)
\begin{equation}
F(q,t)|_{t=0}=B(q)~~~~\mbox{and}~~~~\partial_tF(q,t)_{|_{t=0}}=qA(q).
\end{equation}
It can be checked \cite{SWM} that
\begin{equation}\label{re}
\begin{array}{ll}
F_1(q,t)_{|_{t=0}}=\frac{\sqrt{\pi}~2^{\frac{-\imath
q^2-\check{\mu}_1}{2\check{\mu}_1}}}
{\Gamma(\frac{3}{4}+\imath\frac{q^2}{4\check{\mu}_1})}, &~~~~
\partial_tF_1(q,t)_{|_{t=0}}=\frac{(-1)^{-1/4}\sqrt{\pi}~(-\imath q^2-\check{\mu}_1)
~2^{\frac{-\imath
q^2-\check{\mu}_1}{2\check{\mu}_1}}}{2\sqrt{\check{\mu}_1}~\Gamma(\frac{5}{4}+
\imath\frac{q^2}{4\check{\mu}_1})},\\
F_2(q,t)_{|_{t=0}}= 1,&~~~~\partial_tF_2(q,t)_{|_{t=0}}=0.
\end{array}
\end{equation}
It results from   (\ref{re}) that the solution $F_2(q,t)$ is a
bounded function, so it does not need any redefinition. For $q^2$
big enough, $F_1(q,t)_{|_{t=0}}$ and
$\partial_tF_1(q,t)_{|_{t=0}}$ are found to be (see Eq. (6.1.45)
in \cite{abram})
\begin{equation}
|F_1(q,t)_{|_{t=0}}|\approx
\sqrt[4]{\frac{\check{\mu}_1}{4}}~\frac{\exp{(\frac{\pi}{8\check{\mu}_1}q^2)}}
{\sqrt{q}},~~~~|\partial_tF_1(q,t)_{|_{t=0}}|\approx
\sqrt[4]{\frac{\check{\mu}_1}{4}}~\sqrt{q}\exp{(\frac{\pi}{8\check{\mu}_1}q^2)}.
\end{equation}
Thus, we redefine the  solution $F_1(q,t)$ as follows
\begin{equation}\label{redef}
F_1 (q,t):=
\sqrt{q}\;\exp{(-\frac{\pi}{8\check{\mu}_1}q^2)}\exp{(-i
\check{\mu}_1 t^2/2)}\;H \Big( -\frac{\check{\mu}_1+iq^2 }
   {2 \check{\mu}_1}, (-1)^{1/4}\;\sqrt{\check{\mu}_1} \;t\Big).
\end{equation}
It is clear that  (\ref{redef}) is the solution of (\ref{eqF})
owing to the structure of the equation. Now, one can verify that
\begin{equation}
\begin{array}{ll}
|A_1(q)|=\sqrt[4]{\frac{\check{\mu}_1}{4}},&~~~~|B_1(q)|=\sqrt[4]{\frac{\check{\mu}_1}{4}},
\\A_2(q)=0,&~~~~B_2(q)=1.
\end{array}
\end{equation}
Therefore, we get the result that the functions $\dR\times
[-t_0,t_0] \ni(q,t)\rightarrow F_s (q,t)\in \dC,~~(s=1,2)~$ are
bounded.

\noindent Second, we define the following generalized functions
\begin{equation}\label{hs}
h_s(t,X^1,\ldots,X^{d-1}):=
\int_{\dR^{d-1}}f(q_1,\ldots,q_{d-1})\;F_s(q,t)\prod_k \exp(-i q_k
X^k)\;dq_1\ldots
   dq_{d-1},
\end{equation}
where $~q^2 = q_1^2 + \dots q_{d-1}^2,~$ and where $f\in
L^2(\dR^{d-1})~$. Since $~F_s~$ are bounded, the functions $f F_s
\in L^2([-t_0,t_0]\times \dR^{d-1})$.  Equation (\ref{hs})
includes (\ref{solG}) due to the term $~\exp(-i q_k X^k)$, with
$q_k \in \dR$.

\noindent Finally, we notice that (\ref{hs}) defines the Fourier
transform of $~fF_s$. Therefore, according to the Fourier
transform theory  (see, e.q. \cite{LDP}) the equation (\ref{hs})
defines the mapping
\begin{equation}\label{plan}
 L^2(\dR^{d-1})\ni f \longrightarrow h_s \in  L^2([-t_0,t_0]
 \times \dR^{d-1})=: \tilde{\mathcal{H}}.
\end{equation}
Replacing  $f$ by consecutive elements of a basis in
$L^2(\dR^{d-1})$ leads to an infinite countable  set of vectors in
$\tilde{\mathcal{H}}$. So obtained set of vectors can be
rearranged into a set of independent vectors and further turned
into an orthonormal basis  by making use of the Gram-Schmidt
procedure \cite{EP}. One can show \cite{LDP} that the span of such
an orthonormal basis, $\mathcal{F}$, is dense in
$\tilde{\mathcal{H}}$. The completion of $\mathcal{F}$ defines the
Hilbert space $\mathcal{H}\subseteq \tilde{\mathcal{H}}$.

To illustrate the above construction, let us use the Hilbert space
$L^2(\dR^{d-1}):= \bigotimes_{k=1}^{d-1}L^2_k(\dR)$, where
$L^2_1(\dR)= L^2_2(\dR)= \ldots = L^2_{d-1}(\dR)\equiv L^2(\dR)$.
Let us take a countable infinite set of vectors $f_n \in L^2(\dR)$
defined as
\begin{equation}\label{basis}
    f_n(q):= \frac{1}{\sqrt{2^n n! \sqrt{\pi}}}\;\exp(-q^2/2)\;H_n(q),
    ~~~~~n=0,1,2,\ldots,
\end{equation}
where $H_n(q)$ is the Hermite polynomial. It is proved in
\cite{NIA} that (\ref{basis}) defines an orthonormal basis in
$L^2(\dR)$. The basis (\ref{basis}) can be used to construct a
basis in  $L^2(\dR^{d-1})$. The basis is defined as the set of all
vectors of the form $\bigotimes_{k=1}^{d-1} f_{n_k}(q^k)\in
L^2(\dR^{d-1})$.  Further steps of the procedure leading to the
dense subspace $\mathcal{F}$ are the same as described in the
paragraph including Eq. (\ref{plan}).

It is clear that (\ref{hs}), owing to the above construction,
defines the solution to the equation $\hat{H}_T h_s = 0$.

\goodbreak
\section{Quantum membrane}

The algebra of Hamiltonian constraints describing a membrane
winding around compact dimension of the $\Milne_C$ space is
defined as follows (for notation and more details see \cite{MP9})
\begin{equation}\label{alg1}
    \{ \check{C}_{+}(f),
    \check{C}_{+}(g)\} = \check{C}_{+}(f\acute{g}-\acute{f} g),
\end{equation}
\begin{equation}\label{alg2}
    \{ \check{C}_{-}(f),
    \check{C}_{-}(g)\} = \check{C}_{-}(f\acute{g}-\acute{f} g),
\end{equation}
\begin{equation}\label{alg3}
\{ \check{C}_{+}(f), \check{C}_{-}(g)\} = 0,
\end{equation}
where
\begin{equation}
\check{C}_{\pm}(f)=\int_{-\pi}^\pi \frac{C\pm C_1}{2}f~d\sigma
\end{equation}
and
\begin{equation}
    C:=\frac{1}{2\kappa X^0}\Pi_{\mu} \Pi_{\nu} \eta^{\mu\nu} +
    \frac{\kappa X^0}{2} \;det[ \acute{X}^{\mu}
    \acute{X}^{\nu} \eta_{\mu\nu}]\approx 0,
\end{equation}
\begin{equation}
    C_1 :=  \acute{X}^{\mu} \Pi_{\mu},
\end{equation}
and where the Poisson bracket is defined to be
\begin{equation}\label{bra1}
\{\check{A},\check{B}\}:= \int_{-\pi}^\pi d\sigma\;
\Big(\frac{\partial\check{A}}{\partial X^\mu}
    \frac{\partial\check{B}}{\partial \Pi_\mu}
     - \frac{\partial\check{A}}{\partial \Pi_\mu}
    \frac{\partial\check{B}}{\partial X^\mu}\Big),
\end{equation}
and where $\acute{f}\equiv df/d\sigma$; $(X^\mu)\equiv
(T,X^k)\equiv (T,X^1,\ldots,X^{d-1})$ are the embedding functions
of an uniformly winding membrane in the $\Milne_C$ space; $d+1$ is
dimension of the target space; $\Pi_\mu$ are the canonical momenta
corresponding to $X^\mu$; and `smeared' constraint $\check{A}$ is
defined as
\begin{equation}\label{integ}
    \check{A}:= \int_{-\pi}^\pi d\sigma \;
    f(\sigma)A(X^{\mu}, \Pi_{\mu}),~~~~f \in \{C^\infty [-\pi,\pi]\,|\,
    f^{(n)}(-\pi)=f^{(n)}(\pi)\}.
\end{equation}

Quantization of the algebra (\ref{alg1})-(\ref{alg3}) means
finding its self-adjoint representation in a Hilbert space. It is
clear that (\ref{alg1})-(\ref{alg3}) consists of two independent
subalgebras. To be specific, we first quantize the subalgebra
satisfied by
\begin{equation}\label{cir}
    L_n := \check{C}_{+}(\exp{in\sigma}),~~~~n \in \dZ.
\end{equation}
One may easily verify that
\begin{equation}\label{alg4}
\{L_n,L_m\} = i (m-n) L_{m+n}.
\end{equation}

Quantization of (\ref{alg2}) can be done by analogy.  Merger of
both quantum subalgebras will complete the problem of finding the
representation of the  algebra (\ref{alg1})-(\ref{alg3}).

\subsection{Representations of the constraints}

\subsubsection{Representation based on a  single field}

\textit{Hilbert space}\\
The pre-Hilbert space, $\tilde{\mathcal{H}}$, induced by the space
of fields, $\dS \ni \sigma \rightarrow X(\sigma)$, is defined to
be (see the paper \cite{MP11})
\begin{eqnarray}\label{state}
  \tilde{\mathcal{H}}\ni\Psi[X] &:=& \int \psi(X,\acute{X},\sigma)d\sigma ,
  \\\label{scal1}
  \langle\Psi |\Phi \rangle &:=& {\int\overline{\Psi}[X]\Phi[X][dX]},
\end{eqnarray}
where $\psi(X,\acute{X},\sigma)$ is  such that $ \langle\Psi |\Psi
\rangle<\infty$. The measure $[dX]$ is assumed to be invariant
with respect to $\sigma$ reparametrization. Completion of
$\tilde{\mathcal{H}}$ in the norm induced by (\ref{scal1}) defines
the Hilbert space $\mathcal{H}$.\\
\textit{Representation of  generator}\\
In what follows we find a  representation of (\ref{alg4}). Let us
consider a diffeomorphism on $\dS^1$ of the form $X(\sigma)
\mapsto X(\sigma+\epsilon v(\sigma))$. For a small $\epsilon$ we
have
\begin{eqnarray}\label{xx1}
  X(\sigma+\epsilon
v(\sigma)) &\approx& X(\sigma)+\epsilon v(\sigma)\acute{X}
(\sigma)=: X(\sigma)+\epsilon L_{v}X(\sigma), \\
  \acute{X}(\sigma+\epsilon
v(\sigma))&\approx&
\acute{X}(\sigma)+\epsilon\frac{d}{d\sigma}[v(\sigma)\acute{X}(\sigma)]
= \acute{X}(\sigma)+\epsilon\frac{d}{d\sigma}[L_{v}{X}(\sigma)].
\end{eqnarray}
Now, we define an operator $\hat{L}_{v}$ corresponding to  $L_{v}$
defined by (\ref{xx1}). Since we have
\begin{equation}
\Psi[X(\sigma+\epsilon
v(\sigma))]\approx\Psi[X(\sigma)]+\epsilon\int\Big(
\frac{\partial\psi}{\partial X}L_{v}X+\frac{\partial\psi}{\partial
\acute{X}}\frac{d}{d\sigma}[L_{v}{X}]\Big)d\sigma,
\end{equation}
we set
\begin{equation}\label{l}
\hat{L}_{v}\Psi[X]:=\int \Big(\frac{\partial\psi}{\partial X}L_{v}X+\frac{\partial\psi}{\partial
\acute{X}}\frac{d}{d\sigma}[L_{v}{X}]\Big)d\sigma=\int \Big(\acute{v}\frac{\partial\psi}{\partial
\acute{X}}\acute{X}-\acute{v}\psi-v\frac{\partial\psi}{\partial \sigma}\Big) d\sigma~\in\mathcal{H}.
\end{equation}
One may verify that $\{L_v,L_w \}=L_{(v\acute{w}-\acute{v}w)}$ and
 check that
\begin{equation}\label{alg}
[\hat{L}_{v},\hat{L}_{w}]=\hat{L}_{(v\acute{w} -\acute{v}w)}.
\end{equation}
Next, let us consider the following
\begin{eqnarray}\nonumber
\int\overline{\Psi}[X(\sigma+\epsilon
v(\sigma))]\Phi[X(\sigma)][dX(\sigma)]&=&\int\overline{\Psi}[X(\sigma)]
\Phi[X(\sigma-\epsilon v(\sigma))][dX(\sigma-\epsilon
v(\sigma))]\\ \label{conj}
&=&\int\overline{\Psi}[X(\sigma)]\Phi[X(\sigma-\epsilon
v(\sigma))][dX(\sigma)],
\end{eqnarray}
where we assume that $v(\sigma)$ is a real function and
$\sigma\mapsto \sigma+\epsilon v(\sigma)$ is a diffeomorphism.
Taking derivative with respect to $\epsilon$ of both sides of
(\ref{conj})   and putting $\epsilon=0$ leads to
\begin{equation}\label{algN}
\langle\hat{L}_v\Psi |\Phi \rangle=-\langle\Psi |\hat{L}_v\Phi
\rangle .
\end{equation}
Therefore,  the operator $\hat{L}_n $ defined by the mapping
\begin{equation}\label{Map1}
    L_n \longrightarrow \hat{L}_n :=  i \,\hat{L}_{\exp(i n\sigma)}
\end{equation}
is symmetric on $\mathcal{H}$ and leads to a symmetric
representation of the algebra (\ref{alg4}). It is a self-adjoint
representation if $\hat{L}_n$ are bounded operators \cite{RS}.\\
\textit{Solving the constraint}\\
Since we look for diffeomorphism invariant states,  it is
sufficient to assume that $\psi=\psi(X,\acute{X})$. Let us solve
the equation
\begin{equation}\label{sss}
\hat{L}_n\Psi=0,
\end{equation}
which after making use of (\ref{l}) and integrating by parts reads
\begin{equation}\label{cons1}
\int \acute{(e^{\imath
n\sigma})}[-\psi+\frac{\partial\psi}{\partial
\acute{X}}\acute{X}]~d\sigma=0.
\end{equation}
General solution to (\ref{cons1}) has the form
\begin{equation}\label{cons}
-\psi+\frac{\partial\psi}{\partial \acute{X}}\acute{X}=\sum_{k\neq
-n}a_k e^{\imath k\sigma}~~~~\textrm{for}~n\neq 0 ,
\end{equation}
where $a_k$ are arbitrary constants, and there is no condition for
$n=0$. Our goal is an imposition of all the constraint,  i.e. we
look for  $\Psi :\forall n~\hat{L}_n\Psi=0$. We find that the
intersection of all the kernels defined by (\ref{cons}) is given
by the equation
\begin{equation}\label{conss}
-\psi+\frac{\partial\psi}{\partial \acute{X}}\acute{X}=c,
\end{equation}
where $c$ is an arbitrary constant. It is enough to solve
(\ref{conss})  for $c=0$ and then simply add to the solution any
constant. Since the above equation results from (\ref{cons1}), it
is expected to hold in a more general sense, i.e. in a
distributional sense. It is clear that the space of solutions to
(\ref{conss}) is defined by
\begin{equation}\label{sol}
\psi=\alpha(X)|\acute{X}|+\beta(X)\acute{X}-c,
\end{equation}
where $\alpha$ and $\beta$ are any functions. The first term is a
distribution, the second one can be checked to be trivial, since
\begin{equation}
\int_{\dS^1} \beta(X)\acute{X}~d\sigma=\int_{\dS^1} \beta(X) dX=0
\end{equation}
for a periodic field $X$, and third one is a functional that gives
the same value $2\pi c$ for every field.\\
\textit{Interpretation of solutions}\\
Let us identify special features of the fields $X$ specific to the first term in (\ref{sol})
\begin{eqnarray}\label{dist}
\Psi[X] &=& \int \alpha(X)|\acute{X}|~d\sigma = \int \frac{d}{d\sigma}{[\gamma(X)]}(\tilde{H}(\acute{X})-\tilde{H}(-\acute{X}))~d\sigma
 \\ \nonumber
&=& -\int \gamma(X)2\delta(\acute{X})~d\acute{X} =- \sum_{\textrm{extr}\;  X }2\gamma(X)= \sum_{\textrm{min}\; X }2\gamma(X)-\sum_{
\textrm{max}\; X }2\gamma(X),
\end{eqnarray}
where $d\gamma/dX=\alpha$ and $\tilde{H}$ is the Heaviside
function. Thus, $\Psi$  depends on the values of $\gamma$ at
extrema points of $X$. We have diffeomorphism invariance due to
the implication $(\frac{dX}{d\sigma}=0)\Rightarrow
(\frac{dX}{d\widetilde{\sigma}}=
\frac{d\sigma}{d\widetilde{\sigma}}\frac{dX}{d\sigma}=0)$.\\
\textit{Representation of the algebra}\\
 The mapping (\ref{Map1}) turns (\ref{alg4}) into
\begin{equation}\label{Alg5}
    [\hat{L}_n,\hat{L}_m] = (n - m) \hat{L}_{n+m}.
\end{equation}
 It is clear that our representation is self-adjoint on the space
of solutions to (\ref{sss}), which is  defined by (\ref{dist}), if
$\hat{L}_n$ are bounded operators.

Considerations concerning finding the representation of the
subalgebra (\ref{alg1}) extend directly to the subalgebra
(\ref{alg2}), due to (\ref{alg3}). To construct the representation
of the algebra (\ref{alg1})-(\ref{alg3}), which consists of two
commuting subalgebras, one may use standard techniques
\cite{MP4,EP}. For instance, the representation space of the
algebra may be defined to be either a tensor product or direct sum
of the representations of both subalgebras.

\subsubsection{Representation based on phase space functions}

\textit{Hilbert space}\\
Using the ideas with  the single field case (presented in the
previous subsection) and some ideas from \cite{Thiemann:2004qu},
we construct now the representation of the algebra
(\ref{alg1})-(\ref{alg3}) by making use of the phase space
functions with coordinates $(X^{\mu}, \Pi_{\mu})$, where
$\mu=0,1,\dots, d-1$.

Inspired by \cite{Thiemann:2004qu}, we identify two types of
1-forms on $S^1$, namely $Y^{\lambda}_{\pm}$, which are solutions
to the equation
\begin{eqnarray}\nonumber
    \{C_{\pm}(u), Y^{\lambda}_{\mp}\}=\int\Big( -\frac{d}{d\sigma}\big(2u(T\acute{X}_{\mu}
    \pm\Pi_{\mu})\big)\frac{\delta Y^{\lambda}_{\mp}}{\delta\Pi_{\mu}}-2u(\frac{\Pi^{\mu}}
    {T}\pm \acute{X}^{\mu})\frac{\delta Y^{\lambda}_{\mp}}{\delta
    X^{\mu}}\\ \label{thiemann}
    -u(\frac{\Pi_{\mu}\Pi_{\nu}\eta^{\mu\nu}}{T^2}-\acute{X}^{\mu}\acute{X}^{\nu}
    \eta_{\mu\nu})\frac{\delta
    Y^{\lambda}_{\mp}}{\delta\Pi_0}\Big)d\sigma=0.
\end{eqnarray}
The 1-form $Y^{\mu}_{\pm}$ defines a basis of the plus/minus
sector, respectively. It is clear that an action of $C_{\pm}$ does
not lead outside of a given sector. To be specific, let us first
define the representation for a single sector (for simplicity of
notation we use $Y^{\mu}$ without lower label `plus' or `minus').

As before we propose to include fields $Y^{\mu}(\sigma)$ as well
as their first derivatives $\acute{Y}^{\mu}(\sigma)$ in the
definition of a state
\begin{eqnarray}\label{multistate}
  \mathcal{H}\ni\Psi[\overrightarrow{Y}] &:=& \int \psi(\overrightarrow{Y},
  \acute{\overrightarrow{Y}},\sigma)d\sigma ,\\
  \langle\Psi |\Phi \rangle &:=& \int\overline{\Psi}[\overrightarrow{Y}]
  \Phi[\overrightarrow{Y}][d\overrightarrow{Y}],
\end{eqnarray}
where $\overrightarrow{Y} \equiv (Y^\mu)$, and where
$\psi(\overrightarrow{Y},\acute{\overrightarrow{Y}},\sigma)$ is
any well-behaved function such that $ \langle\Psi |\Psi
\rangle<\infty$.\\
\textit{Solving the constraint}\\
We assume again that
$\psi=\psi(\overrightarrow{Y},\acute{\overrightarrow{Y}})$. Let us
solve the equation
\begin{equation}
\hat{L}_n\Psi[\overrightarrow{Y}]=0,
\end{equation}
which in the case of many fields is a simple extension of
(\ref{cons1}), and reads
\begin{equation}\label{consy}
\int \acute{(e^{\imath
n\sigma})}[-\psi+\frac{\partial\psi}{\partial
\acute{Y}^{\mu}}\acute{Y}^{\mu}]~d\sigma=0.
\end{equation}
By analogy to the single field case we infer that
\begin{equation}
-\psi+\frac{\partial\psi}{\partial
\acute{Y}^{\mu}}\acute{Y}^{\mu}=\sum_{k\neq -n}a_k e^{\imath
k\sigma}~~~~\textrm{for}~n\neq 0
\end{equation}
and again with no condition for $n=0$. Imposing all the
constraints leads to
\begin{equation}
-\psi+\frac{\partial\psi}{\partial
\acute{Y}^{\mu}}\acute{Y}^{\mu}=c .
\end{equation}
One can check that the solutions are of the form
\begin{equation}\label{hilbert}
\psi=\bigg(\sum_i\alpha_i(\overrightarrow{Y})\prod_{\mu}|
\acute{Y}^{\mu}|^{\rho_i^{\mu}}\bigg)^{\frac{1}{\rho}}-c ,
\end{equation}
where $\sum_{\mu}\rho_i^{\mu}=\rho$. This is an expected result
since  the measure
$\sqrt[\rho]{\prod_{\mu}|\acute{Y}^{\mu}|^{\rho^{\mu}}}d\sigma$ is
invariant with respect to $\sigma$-diffeomorphisms.\\
\textit{Interpretation of solutions}\\
Suppose we have a space $V\ni\overrightarrow{Y}$ in which a closed
curve, $\sigma\mapsto Y^{\mu}(\sigma)$, is embedded.  Due to
(\ref{hilbert}) we have a kind of measure in $V$ given by
\begin{equation}\label{measure}
\sqrt[\rho]{\alpha(\overrightarrow{Y})\prod_{\mu}|dY^{\mu}
|^{\rho^{\mu}}}.
\end{equation}
One may say, it is a generalization of the  Riemannian type
metric, since for  $\rho^{\mu}_i =1$ and $\rho=2$ we have
\begin{equation}
\sqrt{g_{\mu\nu}dY^{\mu}dY^{\nu}},
\end{equation}
where $g_{\mu\nu}=g_{\mu\nu}(\overrightarrow{Y})$. In the case,
e.g., $Y^0$ is not a constant field  (\ref{measure}) becomes
\begin{equation}\label{measure1}
\sqrt[\rho]{\alpha(\overrightarrow{Y})\prod_{\mu}|dY^{\mu}|^{\rho^{\mu}}}
=\sqrt[\rho]{\alpha (\overrightarrow{Y})\prod_{\mu\neq
0}\bigg|\frac{dY^{\mu}}{dY^0}\bigg|^{\rho^{\mu}}}|dY^0| =:
\widetilde {\alpha}(Y^0)|dY^0|.
\end{equation}
Thus, it is an extension of the single field metric defined by
(\ref{dist}), which may be rewritten as $\alpha(Y)|dY|$. In this
case however integration (\ref{measure1}) is performed in the
multidimensional space so $\widetilde{\alpha}(Y^0)$  depends on a
particular curve  (not just its end points). In fact, it is a
measure of relative variation of fields, i.e. quantity that is
both gauge-invariant and determines curve uniquely. Two simple
examples of wavefunction for two fields $Y_1$ and $Y_2$ are given
by
\begin{eqnarray}\label{ex1}
  \psi &=& \alpha(Y_1\pm Y_2)|\acute{Y_1}\pm \acute{Y_2}|, \\ \label{ex2}
  \psi &=& \alpha(Y_1 Y_2)|\acute{Y_1}Y_2+Y_1\acute{Y_2}|,
\end{eqnarray}
where in analogy to the single field case, (\ref{ex1}) and
(\ref{ex2}) `measure extrema points' for fields $Y_1\pm Y_2$ and
$Y_1 Y_2$, respectively.

It is clear that finding the representation of the complete
algebra (\ref{alg1})-(\ref{alg3}), may be carried out by analogy
to the single field case by using  standard techniques
\cite{EP,MP4}. For instance, we may define
$\Psi[Y^{\mu}_{+},~Y^{\mu}_{-}]:=\Psi[Y^{\mu}_{+}]\otimes\Psi[Y^{\mu}_{-}]$.

\subsection{Comment}

We conclude that the resolution of the cosmic singularity in the
context of propagation of a membrane in the compactified Milne
space  relies on finding non-trivial quantum states of a membrane
winding uniformly around compact dimension of the $\Milne_C$
space. \textit{Above we have proposed a consistent way to
construct such states.} Finding solution to the equation
(\ref{thiemann}) will complete our quantization procedure, since
it will allow to interpret the states in terms of physical
quantities.

\subsection{Remarks on representations of observables}

In the space of solutions to the constraints there are many types
of measures in the form (\ref{measure}) which may be used to
define a variety of physical Hilbert spaces and representations.
One may associate operators, in physical Hilbert space,  with
homomorphisms $V\mapsto V$. The operators split the Hilbert space
into a set of invariant subspaces, each of which defines a
specific representation. Each subspace is connected with specific
measure and all other measures that are produced by homomorphisms.
For example, the products of the action of homomorphism upon a
metric (of Riemannian manifold) constitute the space of all the
metrics that are equivalent modulo a change of coordinates and all
other metrics that are reductions of the initial metric.

Now, let us consider an infinitesimal homomorphism, $\widehat{O}_u : V \rightarrow V$, of the space $V$ along the vector field
$u=u^{\lambda}(\overrightarrow{Y})\,\partial/\partial Y^{\lambda}$. In what follows we consider an example of representation:

For the special form of (\ref{hilbert}) defined by
\begin{equation}\label{rrep1}
\psi :=
\alpha_{\mu}(\overrightarrow{Y})\acute{Y}^{\mu},~~~~\textrm{or}~~
\Psi[Y]=\int\alpha_{\mu}(\overrightarrow{Y})d{Y}^{\mu},
\end{equation}
we find that \cite{Trautman}
\begin{equation}\label{tra1}
    \widehat{O}_u\bigg(\int\alpha_{\mu}d{Y}^{\mu}\bigg)=\int\big(
u^{\lambda}\alpha_{\mu,\lambda}+u^{\lambda}_{,\mu}\alpha_{\lambda}\big)d{Y}^{\mu}.
\end{equation}


One may verify that the operators $\widehat{O}_u$ and
$\widehat{O}_v$ associated with vector fields $u$ and $v$ satisfy
the algebra
\begin{equation}\label{repal}
[ \widehat{O}_u,\widehat{O}_v ]=\widehat{O}_{[u,v]}
\end{equation}
The representations defined by (\ref{tra1}) and (\ref{repal}) are self-adjoint if the operators are bounded.


%% file: T3.tex
\def\baselinestretch{1}

\nonumchapter{Conclusions}

\def\baselinestretch{1.66}


In this work we propose modelling the early Universe with quantum
elementary objects propagating in a spacetime with
big-crunch/big-bang type singularity. Thus, we assume that quantum
phase of the Universe (describing the cosmological singularity)
includes classical spacetime. It means that our model is not as
radical as, e.g., the loop quantum cosmology models, which are
expressed entirely in terms of self-adjoint operators acting in a
Hilbert space. Our results show that there exist variety of
quantum states of various extended objects that propagate through
the cosmological singularity and thus fulfil the fundamental
criterion of self-consistency: \emph{A physically correct model of
the CS, within the framework of string/M theory, should be able to
describe  propagation of a p-brane, i.e. an elementary object like
a particle, string and membrane, from the pre-singularity to
post-singularity epoch}.\\
\textbf{Summary} \\
We have considered propagation of test particle, string and
membrane across the singularity of the compactified Milne space
\cite{MP3,MP5,MP7,MP6,MP8,MP4,MP2,MP1,MP9,MP11}. Our analysis
includes both classical and quantum level. Now we will sum up our
results and then give proposal for future research.\\
Classical analysis of the motion of particle, string and membrane
led us to identification of two special issues:
\begin{itemize}
    \item If there is no coordinate system covering both cones of the
    $\Milne_C$ space, we are unable to extend geodesics uniquely
    beyond singularity by the very definition. However, the so-called uniformly winding modes of strings and higher dimensional
    objects are insensitive to this issue, since the embedding function related to the compact dimension is integrated
    out and the modes in this special case propagate smoothly
    and uniquely.
    \item If we assign circle topology to the singularity
    and thus obtain a global coordinate system, particle goes
    infinitely many times along the compact dimension while approaching the singularity
    so the particle cannot propagate through it uniquely. However in the case of
    string we have found that all the winding modes propagate
    smoothly and uniquely - it seems that the same holds also for higher dimensional objects.
\end{itemize}

We have quantized the elementary objects by two different methods:
(a) reduced phase space method (see appendix) and (b) the Dirac
method. Mostly we have focused on the latter one. The conclusions
are the following:
\begin{itemize}
    \item A special state of quantum particle propagates
    uniquely through the singularity. Moreover the quantum realm makes it,
    to some extent, natural to join propagation of quantum particle
    across pre-big-bang and post-big-bang epochs into a single Hilbert
    space in an indeterministic manner.
    \item Classical and quantum analysis proves
    that quantum string propagates smoothly and uniquely.
    \item Construction of Hilbert space for membrane
    demonstrates that the existence of quantum membranes in the
    $\Milne_C$ space is possible.
    \item The reduced phase space quantization of particle, presented
    in appendix, allows to speculate about different propagation rules
    and adding new degrees of freedom. It also comes with a different
    concept of the evolution of quantum states of elementary objects and gives an argument supporting deterministic propagation
    of variety of quantum particle's states across the $\Milne_C$.
\end{itemize}

We have found that our model is promising enough to deserve more
detailed examination, which we specify in what follows.\\
\textbf{Next Steps} \\
The CMU is able potentially to provide a complete scenario of the
evolution of the Universe, one in which the DE and DM play a key
role in both the past and the future. However, the CMU is not free
from problems. The most difficult one is the gravitational
instability of the quantum phase. It has been argued \cite{16,17}
that Big-Crunch of the CMU may collapse into a black hole which
would end the evolution of the Universe. In such a case, the CMU
scenario would need to be modified to make sense.

Finding an instability of the quantum phase would mean that the
cosmological singularity should be modelled by another spacetime.
Examination of the (in)stability problem of the CMU scenario is
the natural next step of our research programme.

%% file: T4.tex
\def\baselinestretch{1}

\chapter{Quantization of particle's dynamics by an alternative method}\label{group}

\def\baselinestretch{1.66}


In this chapter, based on \cite{MP7,MP6,MP8}, we will follow an alternative path to a quantum theory of a particle in the $\Milne_C$ space. In
contrast to the Dirac method we will solve the constraint classically and then quantize the remaining, physical, degrees of freedom. We will
make use of the symmetries of the compactified Milne space in order to identify observables. As it has been already indicated the particle's
classical passage through the singularity, though possible, cannot be determined uniquely due to ill-posed Cauchy problem (except very special
states), which has its imprints in Dirac's quantum theory (see section 2.2). Here we will investigate if the alternative quantization sheds new
light on the problem.

\goodbreak
\section{Preliminaries}

The line element in  $\Milne_C$ reads
\begin{equation}\label{line}
    ds^2 =  - dt^2 +t^2 d\theta^2 ,
\end{equation}
where $(t,\theta)\in \dR^1 \times \dS^1$ and we omitted additional
Euclidian dimensions.
\medskip

Solution to the Killing field equations with the metric
(\ref{line}) reads
\begin{equation}\label{Kil} \eta_1 =
\cosh\theta\: \frac{\partial}{\partial t}-
\frac{\sinh\theta}{t}\:\frac{\partial}{\partial \theta},~~~~\eta_2
= \sinh\theta \:\frac{\partial}{\partial t}
-\frac{\cosh\theta}{t}\:\frac{\partial}{\partial
\theta},~~~~\eta_3 = \frac{\partial}{\partial \theta}.
\end{equation}

One may easily verify that the Killing vectors (\ref{Kil}) satisfy
the algebra \begin{equation}\label{com1}
    [\eta_1 , \eta_2]=0,~~~~~[\eta_3 , \eta_2]=\eta_1,~~~~~[\eta_3 ,
    \eta_1]=\eta_2,
\end{equation}
which is the $iso(1,1)$ Lie algebra \cite{Vil}.   The algebra
(\ref{com1}) is well defined locally everywhere in the $\Milne_C$
space with exception of the singularity $t=0$.

\medskip

It is commonly known that  Killing vectors of a spacetime may be
used to find dynamical integrals of a particle, i.e. quantities
which do not change during the motion of a point mass. In our case
there exist three dynamical integrals and they can be determined
as follows
\begin{equation}\label{in1} I_1 := \Pi_t\:\eta_1^t +
\Pi_\theta\:\eta_1^\theta = \Pi_t\cosh\theta -
\Pi_\theta\frac{\sinh\theta}{T}, \end{equation}
\begin{equation}\label{in2} I_2 := \Pi_t\:\eta_2^t +
\Pi_\theta\:\eta_2^\theta = \Pi_t\sinh\theta -
\Pi_\theta\frac{\cosh\theta}{T}, \end{equation}
\begin{equation}\label{in3} I_3 := \Pi_t\:\eta_3^t +
\Pi_\theta\:\eta_3^\theta =   \Pi_\theta ,
\end{equation}
where $\eta_a^T$ and  $\eta_a^\theta$ are components of the
Killing vectors $\eta_a ~(a=1,2,3)$ and $\Pi_t$, $\Pi_{\theta}$
were defined below eq. (\ref{conP}). Making use of
(\ref{in1})-(\ref{in3}) we may rewrite the constraint (\ref{con1})
in the form
\begin{equation}\label{con2}
    \Phi = I_2^2 -  I_1^2 + m^2 = 0.
\end{equation}

For further analysis we introduce the phase space. It is defined
to be the space of all particle geodesics. To describe a geodesic
uniquely one may use two independent dynamical integrals. In case
only one part of the Milne space is available for particle
dynamics, for example with $t<0$, the phase space, $\Gamma$, could
be defined as
\begin{equation}\label{phase}  \Gamma =
\{(I_1,I_2,I_3)~|~I_2^2 - I_1^2 + m^2 =0, \: I_3 = p_\theta \}.
\end{equation}
For the choice  (\ref{phase}) the phase space may be parameterized
by two variables $\sigma$ and $p_\sigma$ in the following way
\begin{equation}\label{par} I_1 = m\:\cosh\sigma,~~~~~I_2 = m\:
\sinh\sigma,~~~~~I_3 = p_{\sigma} .
\end{equation} One can easily check that
\begin{equation}\label{coom}
\{I_1,I_2\}=0,~~~~~\{I_3,I_2\}=I_1,~~~~~\{I_3,I_1\}=I_2,
\end{equation}
where the Poisson bracket is defined as
\begin{equation}\label{bra1}
\{\cdot,\cdot\} = \frac{\partial \cdot}{\partial
p_\sigma}\frac{\partial\cdot}{\partial\sigma} -
\frac{\partial\cdot}{\partial \sigma}\frac{\partial\cdot}{\partial
p_\sigma}.
\end{equation} Thus the dynamical integrals (\ref{in1})-(\ref{in3})
and the Killing vectors (\ref{Kil}) satisfy the same algebra.
Using properties of the Poisson bracket we get
\begin{equation}\label{com2} \{\Phi,I_a\} = 0,~~~~~a=1,2,3.
\end{equation}

 We define classical observables to be real functions on phase space
which are: (i) gauge invariant,  (ii) specify  all time-like
geodesics of a particle, and (iii) their algebra corresponds to
the local symmetry of the phase space. It is clear, due to
(\ref{com2}), that all dynamical integrals are gauge invariant.
There exist two  functionally independent combinations of them
which specify all  time-like geodesics. We use them to represent
particle observables (one may verify that they are gauge
invariant).
\smallskip

Let us denote by $\sing_\down$ the part of spacetime $\Milne_C$
with $t<0$, the big-crunch/big-bang singularity by $\sing$, and
the part of $\Milne_C$ with $t>0$  by   $\sing_\up$.

By definition, a test particle with constant mass does not modify
a background spacetime. Hence, we postulate that a particle
arriving at the singularity $\sing$ from $\sing_\down$ is
`annihilated' at $\sing$ and next, `created' into $\sing_\up$.
There are four interesting cases of propagation  depending on the
way a particle may go across $\sing$. In each case the propagation
must be consistent with the constraint equation (\ref{con2}). At
$\sing$ both $I_1$ and $I_2$ are not well defined.

\subsubsection{Specification of phase space and observables based
on continuous symmetries}

In this subsection we consider the following propagation:
 particle following spiral geodesics winding clockwise the cone
$\sing_\down $ continues to move along clockwise spirals in
$\sing_\up$ (the same concerns propagation along anticlockwise
spirals). Obviously, for $\Pi_\theta =0$ particle trajectories are
just straight lines both in $\sing_\down$ and $\sing_\up$. Apart
from this we take into account the rotational invariance (with
respect to the axis which coincides with the $y^0$-axis of 3d
Minkowski frame defining (\ref{emb})) of the space of particle
trajectories which occur independently in $\sing_\down$ and
$\sing_\up$.

The set of all particle trajectories can be  determined by two
parameters $\:(c_1,c_2)\in \dR^1 \times [0,2\pi[$. Thus, the phase
space $\Gamma_\down$ of a particle in $\sing_\down$ has topology
$\dR^1 \times \dS^1$. The transition of a particle across $\sing$
makes the dynamics in $\sing_\down$ and $\sing_\up$ to be, to some
extent, independent so the phase space $\Gamma_\up$ of a particle
in $\sing_\up$ has also the  $\dR^1 \times \dS^1$ topology.
Therefore, the phase space $\Gamma_C$ of the entire system has the
topology $\dS^1\times\dR^1\times\dS^1$.

Now let us specify the local symmetry of either $\Gamma_\down$ or
$\Gamma_\up$ by defining the Lie algebra of particle observables.
The system has two independent degrees of freedom represented by
the observables $c_1$ and $c_2$. Equation (\ref{sol}) tells us
that $c_2$ has interpretation of position coordinate, whereas
$c_1$ plays the role of momentum. With such an interpretation, it
is natural to postulate the following Lie algebra for either
$\Gamma_\down$ or $\Gamma_\up$.
\begin{equation}\label{com3} \{c_1,c_2\}=1,~~~~~~ \{\cdot,\cdot\}
:= \frac{\partial\cdot}{\partial c_1}\frac{\partial\cdot}{\partial
c_2} - \frac{\partial\cdot}{\partial
c_2}\frac{\partial\cdot}{\partial c_1}.
\end{equation}

Suppose the observables $c_1$ and $c_2$ describe dynamics in
$\sing_\down$, and let us assume that propagations  in
$\sing_\down$ and $\sing_\up$ are independent. In such case it
would be convenient to introduce  two new observables $c_4$ and
$c_3$ in $\sing_\up$ corresponding to $c_1$ and $c_2$. The Lie
algebra in $\Gamma_C$ would be defined as follows
\begin{equation}\label{com4}
\{c_1,c_2\}=1,~~~~\{c_4,c_3\}=1,~~~~\{c_i,c_j\}=0,~~~~~
\textrm{where}~~i=1,2~~~\textrm{and}~~~j=3,4
\end{equation}
with the Poisson bracket
\begin{equation}\label{com5}
\{\cdot,\cdot\} := \frac{\partial\cdot}{\partial
c_1}\frac{\partial\cdot}{\partial c_2} +
\frac{\partial\cdot}{\partial c_4}\frac{\partial\cdot}{\partial
c_3} - \frac{\partial\cdot}{\partial
c_2}\frac{\partial\cdot}{\partial c_1} -
\frac{\partial\cdot}{\partial c_3}\frac{\partial\cdot}{\partial
c_4}.
\end{equation} But from the discussion above it results that
$\Gamma_C$ has only three independent variables.  We can encode
this property modifying (\ref{com4}) and (\ref{com5}) by the
condition $c_4= c_1$. Finally, we get \begin{equation}\label{com6}
\{c_1,c_2\}=1,~~~~~ \{c_1,c_3\}=1,~~~~ \{c_2,c_3\}=0,
\end{equation} with the Poisson bracket
\begin{equation}\label{com7} \{\cdot,\cdot\} =
\frac{\partial\cdot}{\partial c_1}\frac{\partial\cdot}{\partial
c_2} + \frac{\partial\cdot}{\partial
c_1}\frac{\partial\cdot}{\partial c_3} -
\frac{\partial\cdot}{\partial c_2}\frac{\partial\cdot}{\partial
c_1} - \frac{\partial\cdot}{\partial
c_3}\frac{\partial\cdot}{\partial c_1}.
\end{equation}

The  type of propagation we have considered so far is consistent
with the local isometry (i.e., continuous symmetry) of the
compactified Milne space, in both cones independently. In the next
subsection we increase respected symmetries to include the space
inversion (i.e., discrete symmetry).

\subsubsection{Specification based on continuous and discrete
symmetries}

We take into account (as in  case considered in the previous
subsection) that $\sing_\down$ and $\sing_\up$ have the (clockwise
and anticlockwise) rotational symmetry quite independently. Apart
from this we assume that the singularity $\sing$ may `change' the
clockwise type geodesics into anticlockwise ones, and
\textit{vice-versa}.  From mathematical point of view such case is
allowed because at $\sing$ the space dimension disappears, thus
$p_\theta$ is not well defined there, so it may have different
signs in $\sing_\down$ and $\sing_\up$. Therefore, the space of
geodesics has reflection type of symmetry independently in
$\sing_\down$ and $\sing_\up$, which is equivalent to the space
inversion separately in $\sing_\down$ and $\sing_\up$. The last
symmetry is of discrete type, so it is not the isometry of the
compactified Milne space. It is clear that the phase space
$\Gamma_C$ has the topology $\dS^1\times\dR^1\times\dS^1\times
\dZ_2$.

Proposed type of propagation of a particle through $\sing$ may be
characterized by the conservation of $|\Pi_\theta|$ (instead of
$\Pi_\theta $  required in  the previous subsection). The
consequence is that now $|c_1|=|c_4|$ (instead of $c_1=c_4$ of the
previous subsection). To obtain the algebra of observables we
propose to put $c_4=\varepsilon c_1$, where $\varepsilon =\pm 1$
is a new descrete variable, into (\ref{com4}) and (\ref{com5}).
Thus the algebra reads
\begin{equation}\label{com8}
\{c_1,c_2\}=1,~~~~~ \{c_1,c_3\}=\varepsilon,~~~~ \{c_2,c_3\}=0,
\end{equation} with
the Poisson bracket
\begin{equation}\label{com9} \{\cdot,\cdot\} =
\frac{\partial\cdot}{\partial c_1}\frac{\partial\cdot}{\partial
c_2} + \varepsilon\frac{\partial\cdot}{\partial
c_1}\frac{\partial\cdot} {\partial c_3} -
\frac{\partial\cdot}{\partial c_2}\frac{\partial\cdot}{\partial
c_1}
 - \varepsilon\frac{\partial\cdot}{\partial c_3}\frac{\partial\cdot}  {\partial c_1}.
 \end{equation}

\subsubsection{The case trajectories in pre- and post-singularity
epochs are independent}

Now, we assume that there is no connection at all between
trajectories in the upper and lower parts of the Milne space. For
instance, spiral type geodesic winding  the cone in $\sing_\down $
may be `turned' by $\sing$ into straight line in  $\sing_\up $,
and \textit{vice-versa}. In addition we propose that $\Pi_\theta$
may equal zero either in $\sing_\down $ or in $\sing_\up $.
Justification for such choices are the same as in the preceding
subsection. Obviously, the present case also includes transitions
of spiral geodesics into spiral ones, and straight line into
straight line geodesics.

It is clear that now  the algebra of observables coincides with
(\ref{com4}) and (\ref{com5}), and the entire phase space
$\Gamma_C$ has  the topology $\Gamma_\down\times\Gamma_\up :=
(\dS^1\times\dR^1)\times(\dR^1\times\dS^1)$.

\subsubsection{The case space of trajectories has reduced form of
rotational invariance}

There is one more case  we would like to consider: it is obtained
by ignoring the rotational invariance of the $\Milne_C$ space
assumed to exist separately in $\sing_\down $ and $\sing_\up $.
Now we assume that the invariance does occur, but in the entire
spacetime. Consequently, the algebra of observables is defined by
Eq. (\ref{com3}).

Such type of symmetry of the space of geodesics appears,  e.g.  in
case of propagation of a particle in two-dimensional one-sheet
hyperboloid embedded in three-dimensional Minkowski space
\cite{WP} (2d de Sitter space with topology $\dR^1 \times \dS^1$).

\goodbreak
\section{Quantum models}\label{prcqm}

By quantization we mean finding a self-adjoint representation of
the algebra of classical observables\footnote{We do not need the
observables to be well defined globally, which would be required
for finding an unitary representation of the corresponding Lie
group.}. We find that our quantization method  is sufficient for
analysis of evolution of a quantum particle across the vertex of
$\Milne_C$. Such method was used in the papers
\cite{WP,Piechocki:2003hh} dealing with dynamics of a particle in
de Sitter space\footnote{Lifting of self-adjoint representation of
the algebra to the unitary representation of the corresponding Lie
group was possible in case of the spacetime topology $\dR^1 \times
\dS^1$, but could not be done in case of topology $\dR^2$.}.
Applying the same quantization method in both cases enables the
comparison of results.

Before we begin quantization, it is advantageous  to redefine the
algebra (\ref{com6}). It is known (see
\cite{LP2,LP,Gonzalez:1998kj,Kowalski:1998hx,Piechocki:2003hh,Brzezinski:1992gu}
and references therein) that in case canonical variables
$(\pi,\beta)$ have the topology $\dR^1 \times \dS^1$, it is
necessary to replace $\beta$ by $U:=\exp(i\beta)$, and replace the
Poisson bracket
\begin{equation}\label{q1} \{\cdot,\cdot\} =
\frac{\partial\cdot}{\partial \pi}\frac{\partial\cdot} {\partial
\beta} - \frac{\partial\cdot}{\partial
\beta}\frac{\partial\cdot}{\partial \pi}
\end{equation} by the  bracket
\begin{equation}\label{q2} <\cdot,\cdot> :=
\Big(\frac{\partial\cdot}{\partial
\pi}\frac{\partial\cdot}{\partial U}
    - \frac{\partial\cdot}{\partial U}\frac{\partial\cdot}{\partial
    \pi}\Big)U = \{\cdot,\cdot\}U .
\end{equation}
So, in particular one gets $\:<\pi,U>=U~$, instead of
$~\{\pi,\beta\}=1$.

\subsubsection{Quantization corresponding to the continuous
symmetry case}

Applying the  redefinition (\ref{q2}) to the algebra (\ref{com6})
leads to
\begin{equation}\label{q3} \langle c_1,U_2\rangle
=U_2,~~~~~ \langle c_1,U_3\rangle = U_3,~~~~~ \langle
U_2,U_3\rangle = 0 ,
\end{equation}
where $U_2:=\exp(ic_2)$ and $U_3:=\exp(ic_3)$, and where the
algebra multiplication reads
\begin{equation}\label{q4} \langle\cdot,\cdot\rangle :=
\Big(\frac{\partial\cdot}{\partial
c_1}\frac{\partial\cdot}{\partial U_2} - \frac{\partial\cdot}
{\partial U_2}\frac{\partial\cdot}{\partial c_1}\Big)U_2 +
\Big(\frac{\partial\cdot}{\partial
c_1}\frac{\partial\cdot}{\partial U_3} - \frac{\partial\cdot}
{\partial U_3}\frac{\partial\cdot}{\partial c_1}\Big) U_3 .
\end{equation}
One may verify that (\ref{q4}) defines the Lie multiplication.

Now, let us quantize the algebra (\ref{q3}). To begin with, we
define the mappings
\begin{equation}\label{q5}
c_1 \rightarrow \hat{c_1}\psi(\beta)\varphi(\alpha):=
-i\frac{d}{d\beta}\psi(\beta)\varphi(\alpha),
\end{equation}
\begin{equation}\label{qq5}
U_2\rightarrow\hat{U_2}\psi(\beta)\varphi(\alpha)
:=e^{i\beta}\psi(\beta)\varphi(\alpha),~~~~~~  U_3\rightarrow
\hat{U_3}\psi(\beta)\varphi(\alpha):=e^{i\beta}\psi(\beta)
e^{i\alpha}\varphi(\alpha),
\end{equation}
where $0\leq\beta,\alpha <2\pi$. The operators $\hat{c_1},
\hat{U_2}$ and $\hat{U_3}$ act on the space
$\Omega_\lambda\otimes\Ola$, where $\Omega_\lambda,~0\leq\lambda <
2\pi,$ is a dense subspace of $L^2 (\dS^1)$ defined as follows
\begin{equation}\label{q6}
\Omega_\lambda = \{\psi\in L^2(\dS^1)~|~ \psi\in
C^{\infty}[0,2\pi], ~
\psi^{(n)}(2\pi)=e^{i\lambda}\psi^{(n)}(0),~~~ n=0,1,2,\dots\}.
\end{equation}
The space $\Ola$ may be chosen to have  more general form than
$\Omega_\lambda$. For simplicity, we assume that it is defined by
(\ref{q6}) as well. However, we do not require that
$\check{\lambda} = \lambda$, which means that the resulting
representation may be labelled by $\check{\lambda}$ and $\lambda$
independently.

The space $\Omega_\lambda\otimes\Ola$ is dense in
$L^2(\dS^1\otimes\dS^1)$, so the unbounded operator $\hat{c_1}$ is
well defined. The operators $\hat{U_2}$ and $\hat{U_3}$ are well
defined on the entire Hilbert space $L^2(\dS^1\otimes\dS^1)$,
since they are unitary, hence bounded. It is clear that
$\Omega_\lambda\otimes\Ola$ is a common invariant domain for all
three operators (\ref{q5}) and their products.

One may easily verify that
\begin{equation}\label{q7}
[\hat{c_1},\hat{U_2}]=
\widehat{<c_1,U_2>},~~~~~[\hat{c_1},\hat{U_3}]=
\widehat{<c_1,U_3>},~~~~~[\hat{U_2},\hat{U_3}]=
\widehat{<U_2,U_3>},
\end{equation}
($[\cdot,\cdot]$ denotes commutator), which shows that the mapping
defined by (\ref{q5}) and (\ref{qq5}) is a homomorphism.

The operator $\hat{c_1}$ is symmetric on
$\Omega_\lambda\otimes\Ola$, due to the boundary properties of
$\psi\in\Omega_\lambda$. It is straightforward to show that
$\hat{c_1}$ is self-adjoint  by solving the deficiency indices
equation \cite{RS} for the adjoint $\hat{c_1}^*$ of $\hat{c_1}$
(for more details see Appendix A of \cite{WP}).

The space $\Omega_\lambda$ may be spanned by the set of
orthonormal eigenfunctions of the operator $\hat{c_1}$ with
reduced domain from $\Omega_\lambda\otimes\Ola$ to
$\Omega_\lambda$, which are easily found to be
\begin{equation}\label{q8}
f_{m,\lambda}(\beta):= (2\pi)^{-1/2}\exp{i\beta
(m+\lambda/2\pi}),~~~~~~m=0,\pm 1,\pm 2,\ldots
\end{equation}
The space $\Ola$ may be also spanned by the set of functions of
the form (\ref{q8}).

We conclude that the mapping defined by (\ref{q5}) and (\ref{qq5})
leads to the self-adjoint representation of (\ref{q3}).

\subsubsection{Quantization corresponding to the continuous and
discrete symmetries case}

Making use of the method presented in preceding subsection we
redefine the algebra (\ref{com8}) to the form
\begin{equation}\label{q9}
\langle c_1,U_2\rangle =U_2,~~~~~ \langle c_1,U_3\rangle
=\varepsilon U_3, ~~~~~ \langle U_2,U_3\rangle = 0,
\end{equation}
where $\var=\pm 1$. We quantize the algebra (\ref{q9}) by the
mapping
\begin{equation}\label{h5}
c_1 \rightarrow \hat{c_1}\psi(\beta)f_\var\varphi(\alpha):=
-i\frac{d}{d\beta}\psi(\beta)f_\var\varphi(\alpha),~~~~~~
U_2\rightarrow\hat{U_2}\psi(\beta)f_\var\varphi(\alpha)
:=e^{i\beta}\psi(\beta)f_\var\varphi(\alpha),
\end{equation}
\begin{equation}\label{hh5}
U_3\rightarrow
\hat{U_3}\psi(\beta)f_\var\varphi(\alpha):=e^{i\beta\hat{\var}}
e^{i\alpha}\psi(\beta)f_\var\varphi(\alpha):=
e^{i\beta\var}\psi(\beta)f_\var e^{i\alpha}\varphi(\alpha),
\end{equation}
where $\hat{\var}$ is the operator acting on the two-dimensional
Hilbert space $E$ spanned by the eigenstates $f_\var$ defined by
\begin{equation}\label{q11}
\hat{\var}f_\var =\var f_\var.
\end{equation}
It is easy to check that
\begin{equation}\label{q20}
 [\hat{c_1},\hat{U_2}] =\hat{U_2},~~~~~
[\hat{c_1},\hat{U_3}] =\hat{\varepsilon} \hat{U_3}, ~~~~~
[\hat{U_2},\hat{U_3}] = 0 .
\end{equation}
The domain space of operators  (\ref{h5}) and (\ref{hh5}) is
defined to be the space $\Omega_\lambda\otimes E\otimes\Ola~$. It
is evident that $\hat{\varepsilon}$ commutes with all operators,
so the algebra (\ref{q20}) is well defined. It is easy to check
(applying results of preceding subsection) that the representation
is self-adjoint.

\subsubsection{Quantization in case the system consists of two
almost independent parts}

In the last case, the only connection between dynamics in
$\sing_\down$ and $\sing_\up$ is that a particle assumed to exist
in $\sing_\down$, can propagate through the singularity into
$\sing_\up$. It is clear that now  quantization of the system may
be expressed in terms of quantizations done separately in
$\sing_\down$ and $\sing_\up$. To be specific, we carry out the
reasoning for $\sing_\down$:

\noindent The phase space has topology $\Gamma_\down
=\dR^1\times\dS^1$ and the algebra of observables read
\begin{equation}\label{q12}
\langle c_1,U_2\rangle =U_2.
\end{equation}
Quantization of (\ref{q12}) immediately gives
\begin{equation}\label{q13}
    c_1 \rightarrow \hat{c_1}\psi(\beta):=
  -i\frac{d}{d\beta}\psi(\beta),~~~~
  U_2 \rightarrow\hat{U_2}\psi(\beta):=
  e^{i\beta}\psi(\beta),~~~~~~\psi\in\Omega_\lambda,
\end{equation}
which leads to
\begin{equation}\label{q51}
[\hat{c_1},\hat{U_2}]= \widehat{<c_1,U_2>}= \hat{U_2}.
\end{equation}

It is obvious that the same reasoning applies to a particle in
$\sing_\up$.

At this stage we can present quantization of the entire system
having phase space with topology $\Gamma_C :=
\Gamma_\down\times\Gamma_\up$. The algebra of classical
observables reads
\begin{equation}\label{q14}
\langle c_1,U_2\rangle =U_2,~~~~\langle c_4,U_3\rangle =U_3,
\end{equation}
with all other possible Lie brackets equal to zero.

\noindent Quantization of the algebra (\ref{q14})  is defined by
\begin{equation}
c_1 \rightarrow \hat{c_1}\psi(\beta)\varphi(\alpha):=
  -i\frac{d}{d\beta}\psi(\beta)\varphi(\alpha),
~~~~U_2 \rightarrow\hat{U_2}\psi(\beta)\varphi(\alpha):=
e^{i\beta}\psi(\beta)\varphi(\alpha),
\end{equation}
\begin{equation}
c_4 \rightarrow \hat{c_4}\psi(\beta)\varphi(\alpha):=
\psi(\beta)\big(-i\frac{d}{d\alpha}\varphi(\alpha)\big), ~~~~
U_3\rightarrow\hat{U_3}\psi(\beta)\varphi(\alpha):=
  \psi(\beta)e^{i\alpha}\varphi(\alpha),
\end{equation}
where the domain of the operators $\:\hat{c_1}, \hat{c_4},
\hat{U_2}\:$ and $\:\hat{U_3}\:$ is $~\Omega_\lambda\otimes\Ola$.

\noindent It is evident that presented representation is
self-adjoint.

\subsubsection{Time-reversal invariance}

\noindent The system of a test particle in the Milne space is a
non-dissipative one. Thus, its theory should be invariant with
respect to time-reversal transformation $T$. The imposition of
this symmetry upon the quantum system, corresponding to the
classical one enjoying such an invariance, may reduce the
ambiguity of quantization procedure commonly associated with any
quantization method \cite{AST}.

In our case the ambiguity is connected with the freedom in the
choice of $\lambda$. Since $0\leq\lambda<2\pi$, there are infinite
number of unitarily non-equivalent representations for the
algebras of observables considered in the preceding subsections.
One may  reduce this ambiguity  following the method of the
imposition of $T$-invariance used for particle dynamics in de
Sitter's space. However, imposition of the rotational invariance
on the space of trajectories makes the definition of time-reversal
invariance meaningless in cases considered in the first three
subsections. The $T$-invariance may be imposed only on the
dynamics considered in the last subsection. The first step of
quantization for this case is specified by Eqs. (\ref{q12}) and
(\ref{q13}). The imposition of the $T$-invariance upon the system
may be achieved by the requirement of the time-reversal invariance
of the algebra (\ref{q51}). Formally, the algebra is
$\hat{T}$-invariant since
\begin{equation}\label{t1}
    \hat{T}\hat{c_1}\hat{T}^{-1} = -
    \hat{c_1},~~~~~\hat{T}\hat{U_2}\hat{T}^{-1}=\hat{U_2}^{-1},
\end{equation}
where $\hat{T}$ denotes an anti-unitary operator corresponding to
the transformation $T$. The first equation in (\ref{t1}) results
from the correspondence principle between classical and quantum
physics, because $c_1$ has interpretation of momentum of a
particle. The assumed form of $\hat{U_2}$ and  anti-unitarity of
$\hat{T}$ lead to the second equation in (\ref{t1}). The formal
reasoning at the level of operators should be completed by the
corresponding one at the level of the domain space
$\Omega_\lambda$ of the algebra (\ref{q51}). Following
step-by-step the method of the imposition of the $T$-invariance
upon dynamics of a test particle in de Sitter's space, presented
in Sec.(4.3) of \cite{Piechocki:2003hh}, leads to the result that
the range of the parameter $\lambda$ must be restricted to the two
values: $\lambda=0$ and $\lambda=\pi$.

Now, let us take into account that quantum theory is expected to
be more fundamental than its classical counterpart (if the latter
exists). In the context of the time-reversal invariance it means
that $\hat{T}$-invariance may be treated to be more fundamental
than $T$-invariance. Applying this idea to quantum particle in the
$\Milne_C$ space, we may ignore the lack of $T$-invariance of
classical dynamics considered in the first three subsections. For
these cases we propose to mean by the time-reversal invariance the
$\hat{T}$-invariance only. It may be realized  by the requirement
of $\hat{T}$-invariance of the corresponding algebras. For
instance, the algebra (\ref{q20}) is formally $\hat{T}$-invariant
if the observables transform as follows
\begin{equation}\label{t2}
\hat{T}\hat{c_1}\hat{T}^{-1} =-\hat{c_1},
~~~~~\hat{T}\hat{U_2}\hat{T}^{-1}=\hat{U_2}^{-1},
~~~~~\hat{T}\hat{U_3}\hat{T}^{-1}=\hat{U_3}^{-1},
~~~~\hat{T}\hat{\var}\hat{T}^{-1} =\hat{\var}~.
\end{equation}
We require the first equation of (\ref{t2}) to hold. All other
equations in (\ref{t2}) result from the functional forms of
$\hat{U_2}$, $\hat{U_3}$ and $\hat{\var}$, and the anti-unitarity
of $\hat{T}$. These analysis should be completed by the
corresponding one at the level of the the domain space
$\Omega_\lambda\otimes E\otimes\Ola~$ of the algebra (\ref{q20}),
but we do not enter into such details.

The imposition of $\hat{T}$-invariance not only meets the
expectation that a system with no dissipation of energy should
have this property, but also helps to reduce the quantization
ambiguity as it was demonstrated in the simplest case (It is clear
that three other cases enjoy this reduction too.).

\goodbreak
\section{Comment}

In short, what we have proposed above is getting rid of
indeterminacy in passage through singularity by
\begin{enumerate}
    \item Solving constraint classically and thus loosing the
    concept of evolution.
    \item Introducing new degrees of freedom of quantum particle, so it 'knows' its destiny before reaching the singularity.
\end{enumerate}
The new degrees of freedom may seem to be introduced in an
arbitrary way since they came from randomly picked symmetries of
the orbifold, which are connected with the singularity, when we
assign point topology to it, i.e.:
\begin{enumerate}
    \item One can rotate the cones independently.
    \item One can inverse $\theta\mapsto -\theta$ independently in
    the both cones.
    \item The Cauchy problem is ill defined so one may actually join any two
    geodesics at the singularity.
\end{enumerate}

These new degrees of freedom are somehow hidden, at least to the
extent we can 'see' the physical world. It would be interesting to
consider a model of interactions between these new degrees of
freedom but at the present level of understanding the physics of
singularity it seems to be far too speculative and definitely
beyond the scope of this work.

\goodbreak
\section{A new criterion and the problem of time}

As it was pointed out already, even the circle topology
singularity produces a non-trivial obstacle in a way to extend
uniquely a geodesic. The reason is that a particle winds around
the compact dimension infinitely many times before it reaches the
singularity. But one may still ask if a quantum state propagating
across the singularity can be extended beyond it uniquely. We
already tried to answer this question in case of a particle in the
Dirac method in chapter 2. In what follows we propose another
approach.

We found in (\ref{sol}) that
\begin{equation}\label{theta}
\Theta(T)= -\textrm{arsinh}
\bigg(\frac{c_{1}}{mT}\bigg)+c_2,~~~~c_1\in\dR ,~~0 \leq c_2
<2\pi.
\end{equation}
As it was shown in the paper \cite{MP6}, $c_1$ and $c_2$ satisfy
the algebra: $\{c_1,c_2\}=1$. We quantize them according to the
section (\ref{prcqm}), i.e. we replace $c_2$ by $U_2:=\exp(ic_2)$
and assign quantum operators:
\begin{equation}
c_1 \rightarrow \hat{c_1}\psi(\beta):=
-i\frac{d}{d\beta}\psi(\beta),
\end{equation}
\begin{equation}
U_2\rightarrow\hat{U_2}\psi(\beta) :=e^{i\beta}\psi(\beta),
\end{equation}
where $0\leq\beta <2\pi$.

Now inspired by ideas presented in \cite{MP10}, we treat $T$ in
(\ref{theta}) as a classical evolution parameter, which enumerates
an ordered family of operators $\hat{\Theta}$, which comes from
substituting in (\ref{theta}) $c_1$ and $c_2$ with $\hat{c_1}$ and
$\hat{U_2}$, respectively. But for the sake of simplicity, let us
consider the following family of self-adjoint operators:
\begin{equation}
\hat{\Theta}(T)=-\textrm{arsinh} \big(\frac{\hat{c}_1}{mT}\big)
\end{equation}
Now let us study the particle approaching the singularity, i.e.
the limit
\begin{equation}\label{limit}
\lim_{T\rightarrow 0^{\pm}}\langle {\theta}\rangle=
\lim_{T\rightarrow 0^{\pm}}\langle
\psi(\beta)|~{{\hat{\Theta}}}\psi(\beta)\rangle .
\end{equation}
First let us express a general state with the eigenvectors of
$\hat{c_1}$ given in (\ref{q8}):
\begin{equation}\label{spec}
    \psi(\beta)=\sum_{m\in~\mathbb{Z}}a_mf_{m,\lambda}(\beta)
\end{equation}
Now we observe that:
\begin{equation}
   \textrm{arsinh}
\big(\frac{{c}_1}{mT}\big)\approx
\textrm{sgn}(\frac{{c}_1}{T})\ln\big|{\frac{2{c}_1}{mT}}\big|,~~~~\textrm{for~}T\ll
\frac{{c}_1}{m}.
\end{equation}
and from this and (\ref{spec}) we conclude that for all finite
combinations of $f_{m,\lambda}(\beta)$ such that:
\begin{equation}\label{qs}
    \sum \textrm{sgn}(m_j+\lambda/2\pi)|a_{m_j}|^2=0
\end{equation}
the limit (\ref{limit}) exists and reads:
\begin{equation}
\pm\ln{\bigg(\prod_{m_j}|m_j+\lambda/2\pi|^{~\textrm{sgn}(m_j+\lambda/2\pi)|a_{m_j}|^2}\bigg)}.
\end{equation}
So we have learnt that though a classical particle in the limit $T\rightarrow 0$ winds the compact dimension infinitely many times, in quantum
theory there exist such mixtures of states, constrained by (\ref{qs}), for which the limit value of $\langle\theta\rangle$ exists. This
observation may be used to extend the quantum states uniquely beyond the singularity. This seems to be more natural then models constructed in
the previous sections of the appendix, since one does not introduce any new degrees of freedom but rather reduces the Hilbert space to vectors,
which are well-behaving functions of time.
